\pdfoutput=1
\begingroup\expandafter\expandafter\expandafter\endgroup
\expandafter\ifx\csname pdfsuppresswarningpagegroup\endcsname\relax
\else
\fi

\documentclass[preprint,review,11pt]{elsarticle}

\usepackage[utf8]{inputenc}

\usepackage[margin=1in]{geometry}

\usepackage[hyphens]{url}
\biboptions{sort&compress, square, comma}
\usepackage[breaklinks=true, linkcolor=blue, citecolor=blue, colorlinks=true]{hyperref}

\usepackage{graphicx}
\usepackage{caption}
\usepackage{subcaption}

\usepackage{booktabs,multirow}
\usepackage[version=3]{mhchem} 
\usepackage{latexsym,amsmath,amssymb}

\usepackage{mathtools}
\usepackage{tablefootnote}

\usepackage[T1]{fontenc}
\usepackage[english]{babel}
\usepackage{csquotes}
\usepackage{textcomp}

\usepackage{soul}
\usepackage[usenames, dvipsnames]{xcolor}

\usepackage{xr}
\usepackage{algorithm}
\usepackage[noend]{algpseudocode}
\makeatletter
\let\OldStatex\Statex
\renewcommand{\Statex}[1][3]{%
  \setlength\@tempdima{\algorithmicindent}%
  \OldStatex\hskip\dimexpr#1\@tempdima\relax}

\usepackage[binary-units]{siunitx}
\DeclareSIUnit\atm{atm}

\graphicspath{{./figures/}}

\usepackage[textsize=small,textwidth=2.25cm]{todonotes}

\journal{Combustion and Flame}

\begin{document}

\begin{frontmatter}

\title{An investigation of GPU-based stiff chemical kinetics integration methods}

\author[uconn]{Nicholas~J.\ Curtis\corref{cor1}}
\ead{nicholas.curtis@uconn.edu}
\author[osu]{Kyle~E.\ Niemeyer}
\author[uconn]{Chih-Jen Sung}

\address[uconn]{Department of Mechanical Engineering\\
  University of Connecticut, Storrs, CT 06269, USA}
\address[osu]{School of Mechanical, Industrial, and Manufacturing Engineering\\
  Oregon State University, Corvallis, OR 97331, USA}

\cortext[cor1]{Corresponding author}

\begin{abstract}
A fifth-order implicit Runge--Kutta method and two fourth-order exponential integration methods equipped with Krylov subspace approximations were implemented for the GPU and paired with the analytical chemical kinetic Jacobian software \texttt{pyJac}.
The performance of each algorithm was evaluated by integrating thermochemical state data sampled from stochastic partially stirred reactor simulations and compared with the commonly used CPU-based implicit integrator \texttt{CVODE}.
We estimated that the implicit Runge--Kutta method running on a single Tesla C2075 GPU is equivalent to \texttt{CVODE} running on \numrange{12}{38} Intel Xeon E5-4640 v2 CPU cores for integration of a single global integration time step of \SI{e-6}{\second} with hydrogen and methane kinetic models.
In the stiffest case studied---the methane model with a global integration time step of \SI{e-4}{\second}---thread divergence and higher memory traffic significantly decreased GPU performance to the equivalent of \texttt{CVODE} running on approximately three CPU cores.
The exponential integration algorithms performed more slowly than the implicit integrators on both the CPU and GPU.
Thread divergence and memory traffic were identified as the main limiters of GPU integrator performance, and techniques to mitigate these issues were discussed.
Use of a finite-difference Jacobian on the GPU---in place of the analytical Jacobian provided by \texttt{pyJac}---greatly decreased integrator performance due to thread divergence, resulting in maximum slowdowns of \SIrange{7.11}{240.96}{$\times$}; in comparison, the corresponding slowdowns on the CPU were just \SIrange{1.39}{2.61}{$\times$}, underscoring the importance of use of an analytical Jacobian for efficient GPU integration.
Finally, future research directions for working towards enabling realistic chemistry in reactive-flow simulations via GPU\slash SIMT accelerated stiff chemical kinetics integration were identified.
\end{abstract}


\end{frontmatter}

\clearpage

\section{Introduction}
\label{sec:Intro}

The need for accurate chemical kinetic models in predictive reactive-flow simulations has driven the development of detailed oxidation models for hydrocarbon fuels relevant to transportation and energy generation applications.
At the same time, growing understanding of hydrocarbon oxidation processes resulted in orders of magnitude increases in model size and complexity.
Contemporary detailed chemical kinetic models relevant to jet fuel~\cite{Naik2011434}, diesel~\cite{Sarathy:2011kx}, gasoline~\cite{Mehl:2011jn}, and biodiesel~\cite{Herbinet:2010gu} surrogates consist of hundreds to thousands of species with potentially tens of thousands of reactions.
Furthermore, kinetic models for large hydrocarbon fuels tend to exhibit high stiffness that requires implicit integration algorithms for practical solution.

Reactive-flow modeling codes commonly rely on high-order implicit integration techniques to solve the stiff governing equations posed by chemical kinetic models.
The cost of these algorithms scales at best quadratically---and at worst cubically---with the number of species in a model~\cite{Lu:2009gh}, due to repeated evaluation and factorization of the chemical kinetic Jacobian matrix to solve the associated nonlinear algebraic equations through iterative solutions of linear systems of equations.
Several recent studies~\cite{Huang20091814,Bottone2012,Moiz2016123} demonstrated that using even modestly sized chemical kinetic models can incur severe computation cost for realistic reactive-flow simulations.
For example, a single high-resolution Large Eddy Simulation (LES) realization of a diesel spray---using up to 22 million grid cells with a 54-species \emph{n}-dodecane model---for {\SI{2}{\milli\second}} after start of injection with the common implicit CVODE solver~\cite{cvode:2.8.2} took {\num{48000}} CPU core hours and up to {\num{20}} days of wall clock time~\cite{Moiz2016123}.
Lu and Law~\cite{Lu:2009gh} extensively reviewed techniques for reducing the cost of using detailed chemical kinetic models; however, significant cost savings can be realized by using an analytical Jacobian formulation, rather than the typical evaluation via finite difference approximations.
This analytical Jacobian approach eliminates numerous chemical source term evaluations, and for a sparse Jacobian (e.g., formulated in terms of species concentrations) the cost of evaluation can drop to a linear dependence on the number of species in the model~\cite{Lu:2009gh}.

In this work, our efforts to accelerate simulations with chemical kinetics focus on improving the integration strategy itself, by developing new algorithms for high-performance hardware accelerators, such as graphics processing units (GPUs) and similar single-instruction multiple-data\slash thread (SIMD\slash SIMT) devices, increasingly available on supercomputing clusters~\cite{nvidia,xsede,orlcf}.
The ultimate goal of the combustion community is to enable use of detailed kinetic models in realistic reactive-flow simulations---potentially via use of GPU-accelerated chemical kinetics.
However, a clear first step is to reduce the cost of realistic reactive-flow simulations with small-to-moderate sized model to the point where they are practical for iterative design purposes.

\subsection{SIMD\slash SIMT architecture}

Central processing unit (CPU) clock speeds increased regularly over the past few decades---commonly known as Moore's Law---however, power consumption and heat dissipation issues slowed this trend recently.
While multicore parallelism increased CPU performance somewhat, recently SIMD\slash SIMT-enabled processors have gained popularity in high-performance computing due to their greatly increased floating operation per second count.
A SIMD instruction utilizes a vector processing unit to execute the same instruction on multiple pieces of data, e.g., performing multiple floating point multiplications concurrently.
In contrast, a SIMT process achieves SIMD parallelism by having many threads execute the same instruction concurrently.
Many different flavors of SIMD\slash SIMT processing exist:
\begin{itemize}
 \item Modern CPUs have vector processing units capable of executing SIMD instructions (e.g., SSE, AVX2)
 \item GPUs feature hundreds to thousands of separate processing units, and utilize the SIMT model
 \item Intel's Xeon Phi co-processor has tens of (hyperthreaded) cores containing wide-vector units designed for SIMD execution, with each core capable of running multiple independent threads
\end{itemize}
Using the SIMD\slash SIMT parallelism model requires extra consideration to accelerate chemical kinetics integration.

\begin{figure}[htbp]
  \centering
  \includegraphics[width=0.5\linewidth]{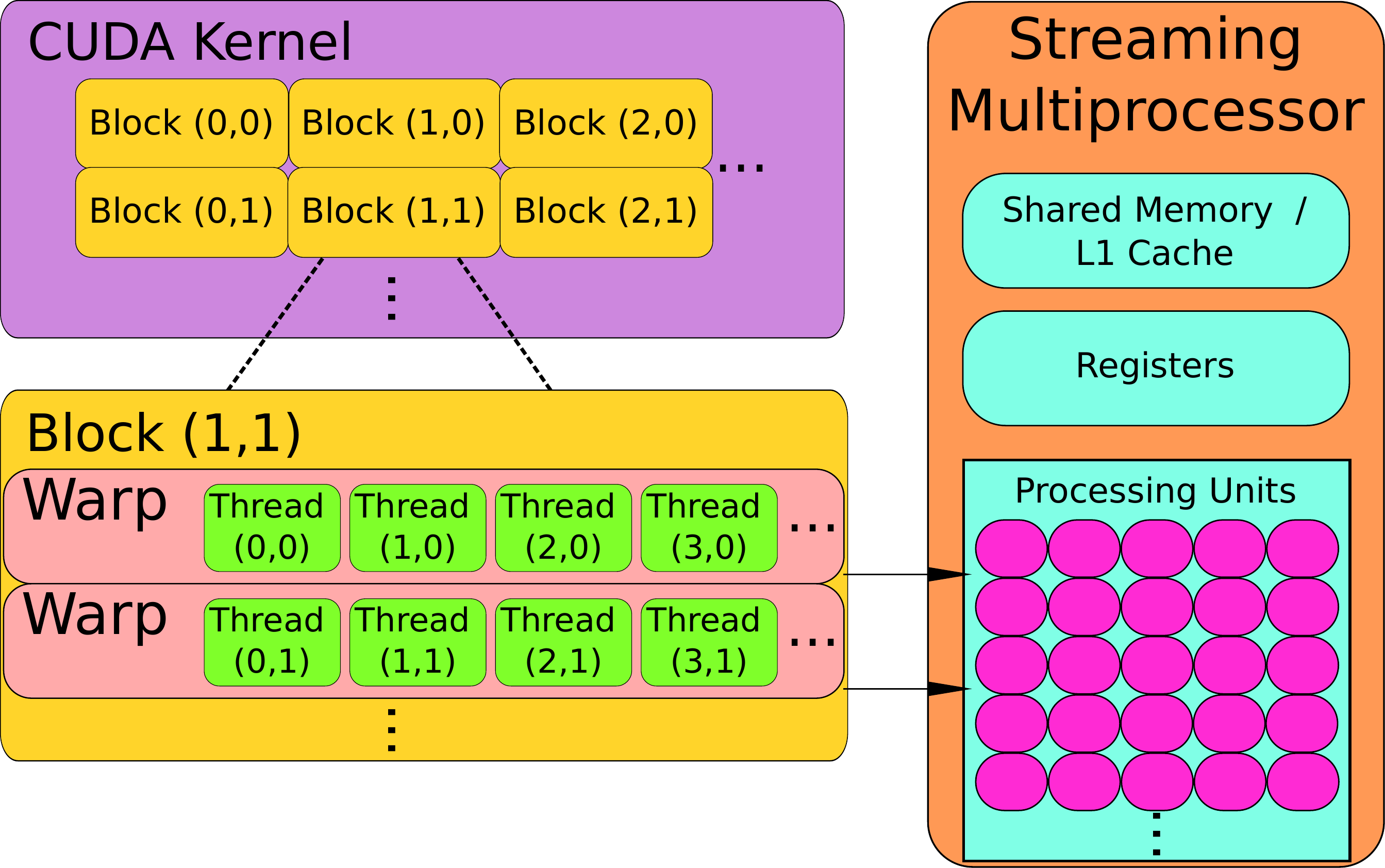}
  \caption{Example of the CUDA SIMT paradigm.
  Program calls (kernels) are split into a grid of blocks, which are in turn composed of threads.
  Threads are grouped in warps (note: warps are typically composed of 32 threads) and executed concurrently on streaming multiprocessors.
  Streaming multiprocessors have registers and L1 cache memory shared between all executing warps.
  Figure file is available under CC-BY~\cite{paperscript:2017}.}
  \label{F:cuda}
\end{figure}

This study used the NVIDIA CUDA framework~\cite{Buck:2008aa,NVIDIA:2015aa}, hence the following discussion will use CUDA terminology; however, the concepts within are widely applicable to SIMT processing.
The basic parallel function call on a GPU, termed a kernel, is broken up into a grid of thread blocks as seen in Fig.~\ref{F:cuda}.
A GPU consists of many streaming multiprocessors (SMs), each of which is assigned one or more thread blocks in the grid.
The SMs further subdivide the blocks into groups of \num{32} threads called warps, which form the fundamental CUDA processing entity.
The resources available on a SM (memory, processing units, registers, etc.) are split between the warps from all the assigned blocks.
The threads in a warp are executed in parallel on CUDA cores (processing units), with multiple warps typically being executed concurrently on a SM.
Thread divergence occurs when the threads in a warp follow different execution paths, e.g., due to if\slash then branching, and is a key performance concern for SIMT processing; in such cases the divergent execution paths must execute in serial.
All threads in a warp are executed even if any thread in the warp is unfinished.
When a divergent path is long and complicated or only a handful of threads in a warp require its execution, significant computational waste may occur as the other threads will be idle for long periods.
A related concept of waste within a SIMD work unit is described by Stone and Niemeyer~\cite{2016arXiv160805794S}.

Furthermore, as compared with a typical CPU, GPUs possess relatively small memory caches and few registers per SM.
These resources are further split between all the blocks\slash warps running on that SM (Fig.~\ref{F:cuda}).
Overuse of these resources can cause slow global memory accesses for data not stored locally in-cache or can even reduce the number of blocks assigned to each SM.
The performance tradeoffs of various CUDA execution patterns are quite involved and beyond the scope of this work; for more details we refer the interested reader to several works that discussed these topics in depth~\cite{Cruz:2011gc,Brodtkorb:2013hn,Niemeyer:2014hn}.
Instead, we will briefly highlight key considerations for CUDA-based integration of chemical kinetic initial value problems (IVPs).

\subsection{GPU-accelerated chemical kinetics}

The extent of thread cooperation within a CUDA-based chemical kinetic IVP integration algorithm is a key point that shapes much of implementation.
GPU-accelerated chemical kinetic solvers typically follow either a ``per-thread'' pattern~\cite{Niemeyer:2011aa,Stone:2013aa,Niemeyer:2014aa}, in which each individual GPU thread solves a single chemical kinetic IVP, or a ``per-block'' approach~\cite{Stone:2013aa,Sewerin20151375}, in which all the threads in a block cooperate to solve the ordinary differential equations (ODEs) that comprise a single chemical kinetic IVP.
The greatest potential benefit of a per-thread approach is that a much larger number of IVPs can theoretically be solved concurrently; the number of blocks that can be executed concurrently on each SM is usually around eight, whereas typical CUDA launch configurations in this work consist of 64 threads per block, or 512 sets of IVPs solved concurrently per SM.
Unfortunately, the larger amount of parallelism offered by a per-thread approach does not come without drawbacks.
A per-thread approach may also encounter more cache-misses, since the memory available per SM must now be split between many more sets of IVPs.
This results in expensive global memory loads.
The performance of a per-thread approach can also be greatly impacted by thread divergence, because different threads may follow different execution paths within the IVP integration algorithm itself~\cite{Stone:2013aa,Niemeyer:2014aa}.
For example, in a per-thread-based solver each thread in a warp advances its IVP by one internal integration step concurrently, and here on a step failure the thread simply does not update the solution vector at the end of the internal time-step.
If only a handful of threads in a warp require many more internal time-steps than the others, they will force the majority of threads to wait until all threads in the warp have completed the global integration step, wasting computational resources.
Additionally, implicit integration algorithms---which typically have complex branching and evaluation paths---may suffer more from thread divergence when implemented on a per-thread basis than relatively simpler explicit integration techniques~\cite{Stone:2013aa}.
The impact of thread divergence on integrators is typically less severe when following a per-block strategy, since the execution path of each thread is planned by design of the algorithm.
A per-block approach also offers significantly more local cache memory and available registers for solving an IVP, and thus memory access speed and cache size are less of a concern.
However, in our experience, optimizing use of these resources requires significant manual tuning and makes it more difficult to generalize the developed algorithm between different chemical kinetic models---a key feature for potential non-academic applications.
In addition, Stone and Davis~\cite{Stone:2013aa} showed that a per-thread implicit integration algorithm outperforms the per-block implementation of the same algorithm in the best-case scenario (elimination of thread divergence by choice of identical initial conditions).

Various studies in recent years explored the use of high-performance SIMT devices to accelerate (turbulent) reactive-flow simulations.
Spafford et al.~\cite{Spafford:2010aa} investigated a GPU implementation of a completely explicit---and thus well suited for SIMT-acceleration---direct numerical simulation code for compressible turbulent combustion.
Using a Tesla C1060 GPU, an order of magnitude speedup was demonstrated for evaluation of species production rates compared to a sequential CPU implementation on a AMD-Operton processor; evaluating chemical source terms is much simpler than chemical kinetics integration on GPUs.
Shi et al.~\cite{Shi:2011aa} used a Tesla C2050 GPU to evaluate species rates and factorize the Jacobian for the integration of (single) independent kinetics systems, showing order-of-magnitude or greater speedups for large kinetic models over a CPU-based code on a quad-core Intel i7 930 processor which used standard \texttt{CHEMKIN}~\cite{kee1989chemkin} and LAPACK~\cite{Anderson:1999aa} libraries for the same operations; it was not clear how\slash if the CPU code was parallelized.
Niemeyer et al.~\cite{Niemeyer:2011aa} implemented an explicit fourth-order Runge--Kutta integrator for a Tesla C2075 GPU, and found a speedup of nearly two orders of magnitude with a nonstiff hydrogen model when compared with a sequential CPU-code utilizing a single core of an Intel Xeon 2.66 GHz CPU.
In a related work, Shi et al.~\cite{Shi:2012aa} developed a GPU-based stabilized explicit solver on a Tesla C2050 and paired it with a CPU-based implicit solver using a single-core of a quad-core Intel i7 930 that handled integration of the most-stiff chemistry cells in a three-dimensional premixed diesel engine simulation; the hybrid solver was \SIrange{11}{46}{$\times$} faster than the implicit CPU solver.
Le et al.~\cite{Le2013596} implemented GPU versions of two high-order shock-capturing reactive-flow codes on a Tesla C2070, and found a \numrange{30}{50}$\times$ speedup over the baseline CPU version running on a single core of a Intel Xeon X5650.
Stone and Davis~\cite{Stone:2013aa} implemented the implicit VODE~\cite{Brown:1989vl} solver on a Fermi M2050 GPU and achieved an order of magnitude speedup over the baseline CPU version running on a single core of a AMD Opteron 6134 Magny-Cours.
They also showed that GPU-based VODE exhibits significant thread divergence, as expected due to its complicated program flow compared with an explicit integration scheme.
Furthermore, Stone and Davis~\cite{Stone:2013aa} found that a per-thread implementation outperforms a per-block version of the same algorithm for $\sim$\num{e4} independent IVPs or more; the per-block implementation reached its maximum speedup for a smaller number of IVPs ($\sim$\num{e3}).
Niemeyer and Sung~\cite{Niemeyer:2014aa} demonstrated an order-of-magnitude speedup for a GPU implementation of a stabilized explicit second-order Runge--Kutta--Chebyshev algorithm on a Tesla C2075 over a CPU implementation of VODE on a six-core Intel X5650 for moderately stiff chemical kinetics.
They also investigated levels of thread divergence due to differing integrator time-step sizes, and found that it negatively impacts overall performance for dissimilar IVP initial conditions in a thread-block.
Sewerin and Rigopoulos~\cite{Sewerin20151375} implemented a three-stage\slash fifth-order implicit Runge--Kutta GPU method~\cite{wanner1991solving} on a per-block basis for high-end (Nvidia Quadro 6000) and consumer-grade (Nvidia Quadro 600) GPUs, as compared to a standard CPU (two-core, four-thread Intel i5-520M) and a scientific workstation (eight-core, 16-thread Intel Xeon E5-2687W) utilizing a message passing interface for parallelization; the high-end GPU was at best \SI{1.8}{$\times$} slower than the workstation CPU (16 threads), while the consumer level GPU was at best \SI{5.5}{$\times$} slower than the corresponding standard CPU (four threads).

While increasing numbers of studies have explored GPU-based chemical kinetics integration, there remains a clear need to find or develop integration algorithms simultaneously suited for the SIMT parallelism of GPUs (along with similar accelerators) and capable of handling stiffness.
In this work we will investigate GPU implementations of several semi-implicit and implicit integration techniques, as compared with their CPU counterparts and the baseline CPU \texttt{CVODE}~\cite{Hindmarsh:2005hg} algorithm.
Semi-implicit methods do not require solution of non-linear systems via Newton iteration---as required for implicit integration techniques---and instead solve sequences of linear systems~\cite{wanner1991solving}; accordingly these techniques are potentially better suited for SIMT acceleration due to an expected reduction of thread divergence (for a per-thread implementation) compared with implicit methods.

Several groups~\cite{Perini20141180,McNenly2015581} previously suggested so-called matrix-free methods as potential improvements to the expensive linear-system solver required in standard implicit methods.
These methods do not require direct factorization of the Jacobian, but instead use an iterative process to approximate the action of the factorized Jacobian on a vector.
Furthermore, Hochbruck and Lubich~\cite{Hochbruck:1997,Hochbruck:1998} demonstrated that the action of the matrix exponential on a vector obtained using Krylov subspace approximation converges faster than corresponding Krylov methods for the solution of linear equations.
Others explored these semi-implicit exponential methods for applications in stiff chemical systems~\cite{Bisetti:2012jw,falati2011integration} and found them stable for time-step sizes greatly exceeding the typical stability bounds.

Since GPU-based semi-implicit exponential methods have not been demonstrated in the literature, we aim to conduct a systematic investigation to test and compare their performance to other common integration techniques.
Finally, we will study the three-stage\slash fifth-order implicit Runge--Kutta algorithm~\cite{wanner1991solving} investigated by Sewerin and Rigopoulos~\cite{Sewerin20151375} here to determine the impact of increasing stiffness on the algorithm and the performance benefits of using an analytical Jacobian matrix, such as that developed by Niemeyer et al.~\cite{niemeyer_2016_51139,Niemeyer:2015ws,Niemeyer:2016aa}.

Recently, implicit methods improved using adaptive preconditioners have shown promise in reducing integration costs for
large kinetic models, compared with implicit methods based on direct, dense linear algebra~\cite{mcnenly2013adaptive}.
These require use of linear iterative methods in addition to the standard Newton iteration, and thus we expect increased
levels of thread-divergence (and integrator performance penalties) for the per-thread approach used in this work.
However, this area merits future study.

The rest of the paper is structured as follows.
Section~\ref{S:method} lays out the methods and implementation details of the algorithms used here.
Subsequently, Sec.~\ref{S:results} presents and discusses the performance of the algorithms run using a database of partially stirred reactor thermochemical states, with particular focus on the effects of thread divergence and memory traffic.
Further, this work is a starting point to reduce the cost of reactive-flow simulations with realistic chemistry via SIMT-accelerated chemical kinetics evaluation.
Thus, we explore the potential impact of current state-of-the-art GPU-accelerated stiff chemical kinetic evaluation on large-scale reactive-flow simulations in Sec.~\ref{S:results}, while identifying the most promising future directions for GPU\slash SIMT accelerated chemical kinetic integration in Sec.~\ref{S:conclusions}.
The source code used in this work is freely available~\cite{accelerInt:beta}.
\ref{S:supp} discusses the validation and performance data, plotting scripts, and figures used in creation of this paper,
as well as the supplementary material which includes unscaled plots of integrator runtimes and characterizations of the partially stirred reactor data for this work.

\section{Methodology}
\label{S:method}

In this section, we discuss details of the algorithms implemented for the GPU along with third-party software used.
The generation of testing conditions will be discussed briefly, and the developed solvers will be verified for expected order of error.

\subsection{Integration techniques}

\begin{table}[htb]
\centering
\begin{tabular}{@{}l c c@{}}
\toprule
Method & CPU & GPU \\
\midrule
\texttt{CVODE} & $\times$ & \\
\texttt{Radau-IIA} & $\times$ & $\times$ \\
\texttt{exp4} & $\times$ & $\times$ \\
\texttt{exprb43} & $\times$ & $\times$ \\
\bottomrule
\end{tabular}
\caption{The solvers used in this study, and platforms considered for each.}
\label{T:solvers}
\end{table}

We investigated GPU implementations of three integration methods in this work, namely \texttt{Radau\-IIA}~\cite{wanner1991solving}, \texttt{exp4}~\cite{Hochbruck:1998}, and \texttt{exprb43}~\cite{Hockbruck:2009}, comparing them against equivalent CPU versions and a CPU-only implicit algorithm \texttt{CVODE}~\cite{Hindmarsh:2005hg,cvode:2.8.2}.
Table~\ref{T:solvers} lists these solvers and their corresponding platforms.
While we describe important details or changes made in this work, full descriptions of all algorithms may be found in the cited sources.
The \texttt{pyJac} software~\cite{niemeyer_2016_51139,Niemeyer:2015ws,Niemeyer:2016aa} provided subroutines for both chemical source terms and the analytical constant-pressure, mass-fraction-based Jacobian matrix used by CPU- and GPU-based algorithms.
We evaluated the relative performance impact of using a finite-difference Jacobian matrix (as compared with an analytical Jacobian) for both platforms with a first-order finite difference method based on that of \texttt{CVODE}~\cite{Hindmarsh:2005hg}.
\texttt{pyJac} also provided the chemical source terms used by the finite-difference Jacobian in all cases.
We direct readers to our previous work~\cite{Niemeyer:2015ws,Niemeyer:2016aa} for verification and performance assessments of \texttt{pyJac} itself.

First, the \texttt{CVODE} solver~\cite{Hindmarsh:2005hg,cvode:2.8.2} (part of the \texttt{SUNDIALS} suite~\cite{sundials:2.6.2}) provided the baseline performance of a typical CPU-based (maximum of fifth-order) implicit integration technique.
In addition, we developed CPU versions of the methods under investigation for direct comparison to the high-order implicit technique.
These include the three-stage\slash fifth-order implicit Runge--Kutta algorithm~\cite{wanner1991solving} (\texttt{Radau-IIA}), the fourth-order exponential Rosenbrock-like method of Hochbruck et al.~\cite{Hochbruck:1998} (\texttt{exp4}), and the newer fourth-order exponential Rosenbrock method~\cite{Hockbruck:2009} (\texttt{exprb43}).
For the exponential methods, we used the method of rational approximants~\cite{gallopoulos:1992} paired with the Carath\'edothy--Fej\'er method~\cite{trefethen:2006,kyle_niemeyer_2016_44291} to approximate the action of the matrix exponential on a vector, as suggested by Bisetti~\cite{Bisetti:2012jw}.
This technique relied on the external \texttt{FFTW3} library~\cite{frigo2005design,fftw:3.3.4}.
However, unlike the approach of Bisetti~\cite{Bisetti:2012jw}, we developed a custom routine based on the algorithm presented by Stewart~\cite{stewart:1998} to perform LU decomposition of the Hessenberg matrix resulting from the Arnoldi iteration.
Convergence of the Arnoldi iteration algorithm was computed using the second term of the exponential matrix\slash vector product infinite series, as suggested in several works~\cite{Bisetti:2012jw,saad:1992}.
The exponential integrators used a rational approximant of type $\left(10,10\right)$ as suggested by Bisetti~\cite{Bisetti:2012jw}.
To ensure high performance of CPU-based methods, the Intel \texttt{MKL} library version 11.3.2 handled linear algebra (i.e., BLAS\slash LAPACK) operations.
Next, we developed GPU versions of the \texttt{Radau-IIA}, \texttt{exp4}, and \texttt{exprb43} methods.
These follow the same descriptions as the CPU versions, but require specialized implementations of several BLAS and LAPACK methods, mostly related to LU factorization of the Jacobian or Hessenberg matrices.
All GPU routines were developed using the NVIDIA CUDA framework~\cite{Buck:2008aa,NVIDIA:2015aa}, and a block-size of 64 threads (8 blocks per SM) was found to be most efficient for all solvers.
All solvers used adaptive time-stepping techniques; the \texttt{Radau-IIA} and \texttt{CVODE} integrators have built-in adaptive time-stepping, while the exponential methods, \texttt{exp4} and \texttt{exprb43}, used a standard adaptive time-stepping technique~\cite{wanner1991solving}.
The adaptive time stepping procedures of all integrators used absolute and relative tolerances of \SI{e-10} and \SI{e-6}, respectively, throughout the work.
Finally, the Jacobian was reused on a per-thread (per-IVP) basis according to the built-in rules for the implicit methods,
and only recomputed on step failures for the exponential methods.

\subsection{Testing conditions}
\label{S:pasr_conditions}

In order to measure the performance of the integrators for realistic conditions, a database of thermochemical states covering a wide range of temperatures and species mass fractions was generated using a previously developed constant-pressure stochastic partially stirred reactor (PaSR) code~\cite{Niemeyer:2016aa,niemeyer_2016_51139}.
We selected two chemical kinetic models to span the range of model sizes typically used in high-fidelity simulations: the hydrogen model of Burke et al.~\cite{Burke:2011fh} with 13 species and 27 reactions, and the GRI-Mech 3.0 model for methane with 53 species and 325 reactions~\cite{smith_gri-mech_30}.
The PaSR simulations were performed at the conditions listed in Table~\ref{T:pasr_parameters} for 10 residence times to reach a statistical steady state; Niemeyer et al.~\cite{Niemeyer:2016aa} describe the PaSR simulation process in greater detail, which follows approaches used by others~\cite{Chen:1997ta,Pope:1997wu,Ren:2014cd}.
The PaSR particles were initialized using the equilibrium state, and gradually move away from equilibrium conditions due to mixing, inflow, and outflow.
In order to reduce the influence of equilibrium conditions on the solution runtime trends for small numbers of IVPs, the first \num{1000} datapoints were removed from each database; this corresponds to a single pairing time, $\tau_\text{pair}$, the time interval at which selected particles in the reactor are randomly swapped with inflowing particles.
At this point in the simulation, $\sim\SI{80}{\percent}$ of the particles were at or near an equilibrium state, and by the \num{5000}th datapoint only $\sim\SI{20}{\percent}$ of the particles were near equilibrium.
The hydrogen and GRI-Mech 3.0 databases consisted of \num{899900} and \num{449900} total conditions, respectively.
Further characterization of the PaSR conditions used in this work can be found in~\ref{S:supp} and our previous study~\cite{Niemeyer:2016aa}.

\begin{table}[htb]
\centering
\begin{tabular}{@{}l c c@{}}
\toprule
Parameter & \ce{H2}\slash air & \ce{CH4}\slash air \\
\midrule
$\phi$ & \multicolumn{2}{c}{1.0} \\
$T_{\text{in}}$ & \multicolumn{2}{c}{\SIlist{400;600;800}{\kelvin}} \\
$p$ & \multicolumn{2}{c}{\SIlist{1;10;25}{\atm}} \\
$N_p$ & \multicolumn{2}{c}{100} \\
$\tau_{\text{res}}$ & \SI{10}{\milli\second} & \SI{5}{\milli\second} \\
$\tau_{\text{mix}}$ & \SI{1}{\milli\second} & \SI{1}{\milli\second} \\
$\tau_{\text{pair}}$ & \SI{1}{\milli\second} & \SI{1}{\milli\second} \\
\bottomrule
\end{tabular}
\caption{
PaSR parameters used for hydrogen\slash air and methane\slash air premixed combustion cases, where $\phi$ indicates equivalence ratio, $T_{\text{in}}$ is the temperature of the inflowing particles, $p$ is the pressure, $N_p$ is the number of particles in the reactor, $\tau_{\text{res}}$ is the residence time, $\tau_{\text{mix}}$ is the mixing time, and $\tau_{\text{pair}}$ is the pairing time.
}
\label{T:pasr_parameters}
\end{table}

\subsection{Solver verification}

\begin{figure}[htbp]
  \centering
  \includegraphics[width=0.7\linewidth]{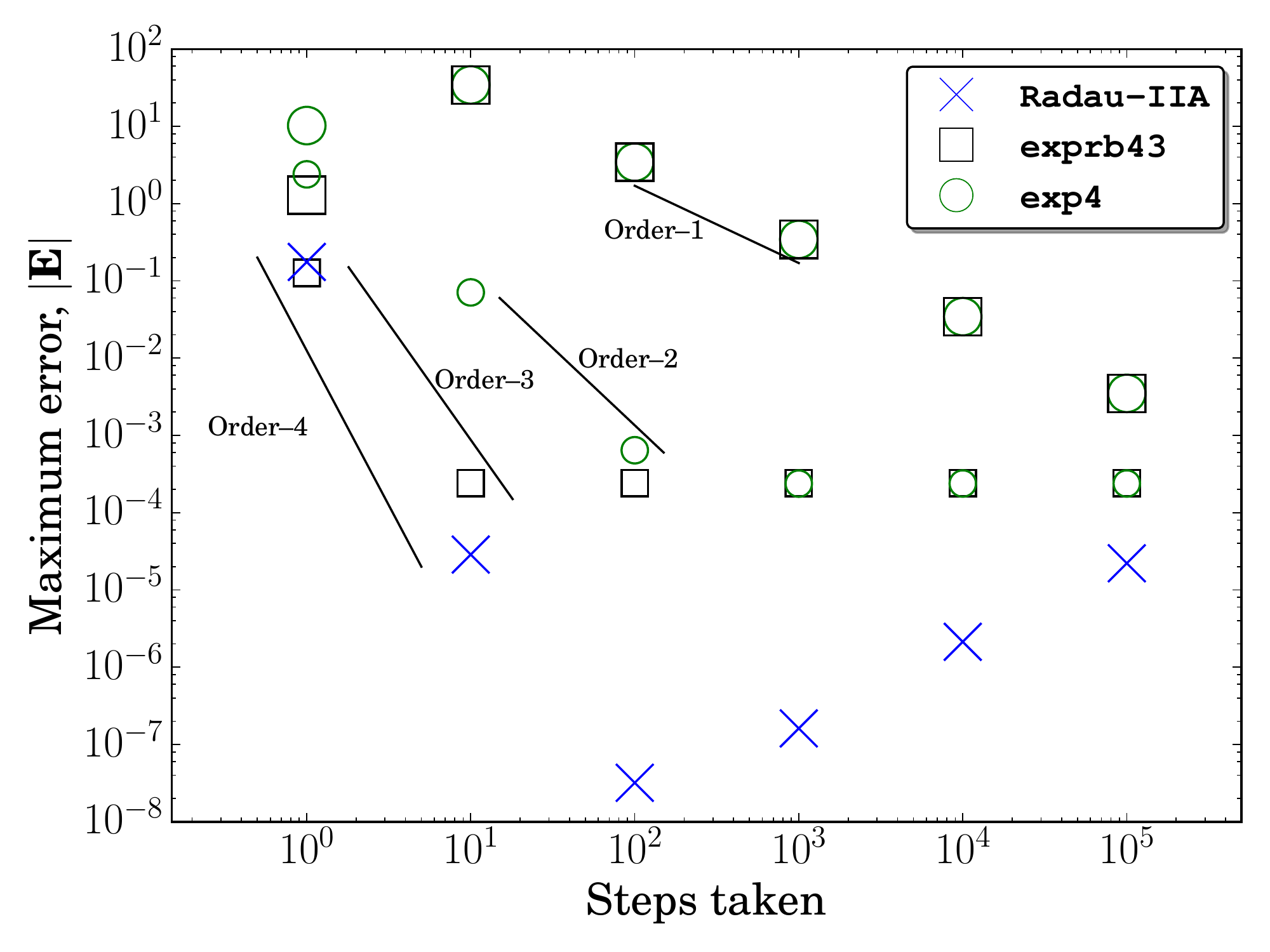}
  \caption{Maximum error of the three CPU solvers as a function of the total number of internal integration steps taken (corresponding to decreasing time-step size).
  Larger square and circle symbols indicate the use of Krylov subspace approximations with the exponential methods, while the smaller symbols indicate the use of ``exact'' Krylov subspaces.
  Data, plotting scripts, and figure file are available under CC-BY~\cite{paperscript:2017}.}
  \label{F:convergence}
 \end{figure}

To investigate the correctness of the developed solvers, the first \num{10000} conditions in the hydrogen database were integrated by each solver using a global time-step size of \SI{e-6}{\second}.
The error for condition $i$ was then determined using the weighted root-mean-square error
\begin{equation}
 E_i(t) = \left\lVert\frac{y_i(t) - \hat{y}_i(t)}{\text{atol} + \hat{y}_i(t) \times \text{rtol}}\right\rVert_2 \;,
\end{equation}
where the $y_i(t)$ is the solution obtained from the various solvers, atol\slash rtol are the absolute\slash relative tolerances, and $\hat{y}_i(t)$ is the ``true'' solution obtained via \texttt{CVODE} using the same global time-step of $\Delta t = \SI{e-6}{\second}$ and absolute\slash relative tolerances of \num{e-20} and \num{e-15}, respectively; note that the more stringent tolerances were used only to obtain the ``true'' solution.
The maximum error over all conditions:
\begin{equation}
 \left\lvert\textbf{E}\right\rvert = \max_{i= 1, \dots, \num{10000}}\{E_i(t)\}
\end{equation}
was then used to measure the error of each solver.
The error measurement used the same tolerances as for the performance testing ($\text{atol} = \num{e-10}$ and $\text{rtol} = \num{e-6}$, respectively).
The constant internal time-step size was then varied from \SIrange{e-6}{e-11}{\second}---corresponding to \numrange[retain-zero-exponent]{e0}{e5} internal integration steps---to measure the convergence rates of the three solvers used in this study.

Figure~\ref{F:convergence} shows the convergence of error for the CPU solvers with decreasing internal time-step size, shown as increasing number of integration steps taken.
The error of the \texttt{Radau-IIA} integrator drops nearly four orders of magnitude when changing from a single internal time step of \SI{e-6}{\second} to ten internal time steps of \SI{e-7}{\second} each, i.e., fourth-order convergence.
Increasing the number of integration steps---by further reducing the internal time-step size---past this point results in one further drop in error (of order $\sim$ 3); however for more than \num{e3} steps the overall error begins to climb due to accumulation of local error.
Since the \texttt{Radau-IIA} solver is nominally fifth-order, it is unclear whether we are observing order reduction due to the stiffness of the problem, use of a numerically obtained ``true'' solution, or an accumulation of local error.
Although a more accurate assessment of convergence order might be achieved through use of a stiff sample problem with an analytical solution---e.g., HIRES~\cite{wanner1991solving} or ROBER \cite{robertson1966solution}---direct validation with the problem at hand was conducted here.

The exponential solvers utilizing an approximate Krylov subspace exhibit larger levels of error in general, with $\left\lvert\textbf{E}\right\rvert \sim \mathcal{O}(1)\text{--}\mathcal{O}(10)$ for a single internal integration step of $\delta t = \SI{e-6}{\second}$.
As the time-step size is decreased, the convergence of the Arnoldi algorithm is affected by the internal integration time-step size (the matrix exponentials and error estimates are scaled by the internal time-step).
To study the effect of the Arnoldi algorithm on error, Fig.~\ref{F:convergence} also presents the error convergence of the exponential integrators with the Krylov approximation error reduced far below the error of the overall method (for larger internal time-steps).
Practically, this was accomplished by detecting when the $n$th Krylov subspace vector approaches zero, a condition known as the ``happy breakdown'' in literature~\cite{datta2010numerical}.
At this limit, the approximate exponential matrix\slash vector product approaches the exact value and thus the Krylov approximation error is relatively small compared to the error of the overall method.
It is clear the that error induced by the ``exact'' Krylov subspace is non-zero however, as both methods reach a minimum error around \num{e2} steps and are unaffected by further step-size decreases, in contrast to the \texttt{Radau-IIA} solver which exhibits increasing error past this point due to local error accumulation.
Figure~\ref{F:convergence} shows that the exponential methods achieve only first-order convergence to the true solution with the approximate Krylov subspace, but both methods converge at higher rates with the ``exact'' Krylov subspace.
The nominal fourth-order convergence of the \texttt{exp4} algorithm is a classical nonstiff order, and thus order reduction is expected for stiff problems~\cite{ANU:7701740,Bisetti:2012jw}; the \texttt{exp4} solver reaches roughly second-order convergence with the ``exact'' Krylov subspace.
The \texttt{exprb43} solver reaches third-order convergence with the ``exact'' Krylov subspace.
Similar to the discussion on the \texttt{Radau-IIA} convergence order, it is difficult to determine whether order reduction has occurred due to problem stiffness, the use of a numerically obtained ``true'' solution, or some combination thereof.
Furthermore, the error of Krylov subspace approximation dominates the error measurement $\lvert\textbf{E}\rvert$.
From Fig.~\ref{F:convergence} we conclude that all three solvers produce reasonably accurate solutions as compared with \texttt{CVODE}.
Additionally, although not shown, the GPU solvers produce identical results.

\section{Results and discussion}
\label{S:results}

We studied the performance of the three integrators by performing constant-pressure, homogeneous reactor simulations with two global integration time-step sizes, $\Delta t = \SI{e-6}{\s}$ and $\Delta t = \SI{e-4}{\s}$, for each integrator. Initial conditions were taken from the PaSR databases described in Sec.~\ref{S:pasr_conditions}.
A larger global time step induces additional stiffness and allows evaluation of the performance of the developed solvers on the same chemical kinetic model with varying levels of stiffness.
In reactive-flow simulations, the chemical integration time-step is typically determined by the flow time-scale and stability requirements determined by the Courant--Friedrichs--Lewy number.
Typical global time-step values of reactive-flow simulations are not always clear in the literature, as adaptive time-stepping is often used, or the global time-step size is simply not reported; our own experience suggests global time-step sizes ranging from \SI{e-7}{\s} to \SI{e-4}{\s}.
The global time-step size used in a given simulation depends highly on the problem and numerical methods, but large-eddy simulations usually require higher time resolution than Reynolds-averaged Navier--Stokes simulations~\cite{Iaccarino:2003147}.
Hence, the global time-step sizes we selected for study represent realistic values used in large-eddy~\cite{Wang20111319,Bulat20133155} and Reynolds-averaged Navier--Stokes~\cite{Ramirez2010,Galloni20091131} simulations.

Runtimes are reported as the average over five runs, where each run started from the same set of PaSR conditions.
All CPU integrators were compiled using \texttt{gcc 4.8.5} (with the compiler options ``\texttt{-O3 -funroll-loops -mtune=native}'') and executed in parallel via OpenMP on four ten-core \SI{2.2}{\giga\hertz} Intel Xeon E5-4640 v2 CPUs with \SI{20}{\mega\byte} of L3 cache memory, installed on an Ace Powerworks PW8027R-TRF+ with a Supermicro X9QR7-TF+/X9QRi-F+ baseboard.
OpenMP was used to parallelize on a per-condition basis; i.e., each individual OpenMP thread was responsible for integrating a single chemical kinetic IVP, rather than cooperating with other OpenMP threads to solve the same.
A six-core \SI{2.67}{\giga\hertz} Intel Xeon X5650 CPU hosted the GPU integrators, which were compiled using \texttt{nvcc 7.5.17} (with compiler options ``\texttt{-arch=sm\_20 -O3 -maxrregcount 63 -{}-ftz=false -{}-prec-div=true -{}-prec-sqrt=true -{}-fmad=false}'') and run on a single NVIDIA Tesla C2075 with \SI{6}{\giga\byte} of global memory.
Reported runtimes for the GPU-based algorithms include time needed for CPU--GPU data transfer before and after each global time step; in addition, the function \texttt{cudaSetDevice()} initialized the GPU before timing to avoid any device initialization delay.
The open-source \texttt{pyJac} software~\cite{niemeyer_2016_51139,Niemeyer:2015ws,Niemeyer:2016aa} produced CPU and GPU custom source-code functions for the chemical source terms and analytical Jacobian matrix evaluation.
Finally, the L1\slash shared-memory cache was set to prefer a larger L1 cache using the \texttt{cudaDeviceSetCacheConfig()} function.

\subsection{Runtime performance}
\label{S:perf}

For all cases in this section, the integrator runtimes are presented as the runtime per IVP solved, for two reasons.
First, saturation of the available computational resources becomes visually apparent (transition from a nearly linear decrease to a flat trend), and second, it allows certain other performance trends (e.g., the effects of thread divergence) to be easily highlighted.
The presentation of the performance data in raw form is also available in the supplementary material for completeness.

\begin{figure}[htbp]
  \centering
  \begin{subfigure}{0.49\textwidth}
      \includegraphics[width=\linewidth]{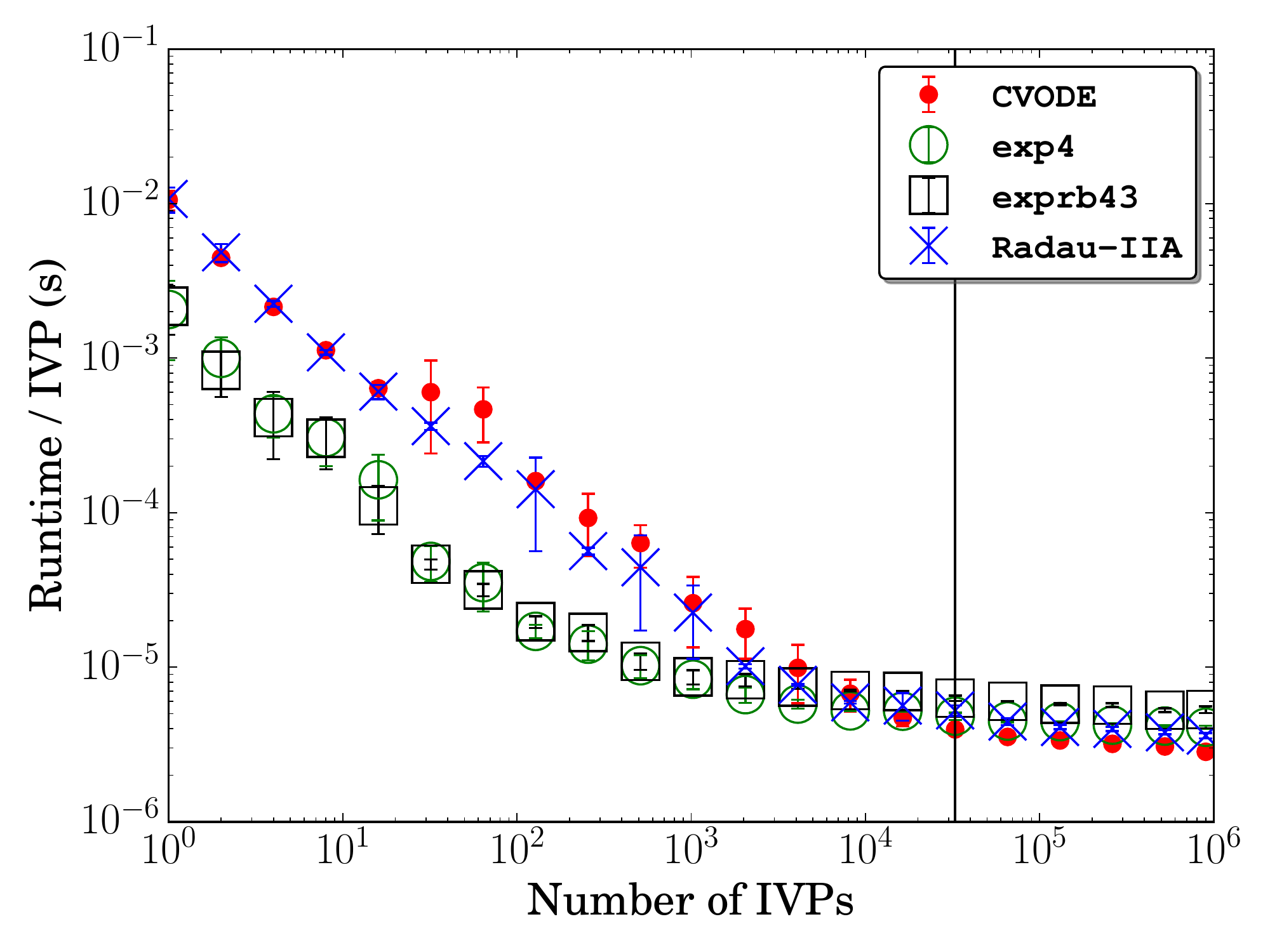}
      \caption{CPU performance results for $\Delta t = \SI{e-6}{\second}$}
      \label{F:h2_cpu_perf_small}
  \end{subfigure}
  \begin{subfigure}{0.49\textwidth}
      \includegraphics[width=\linewidth]{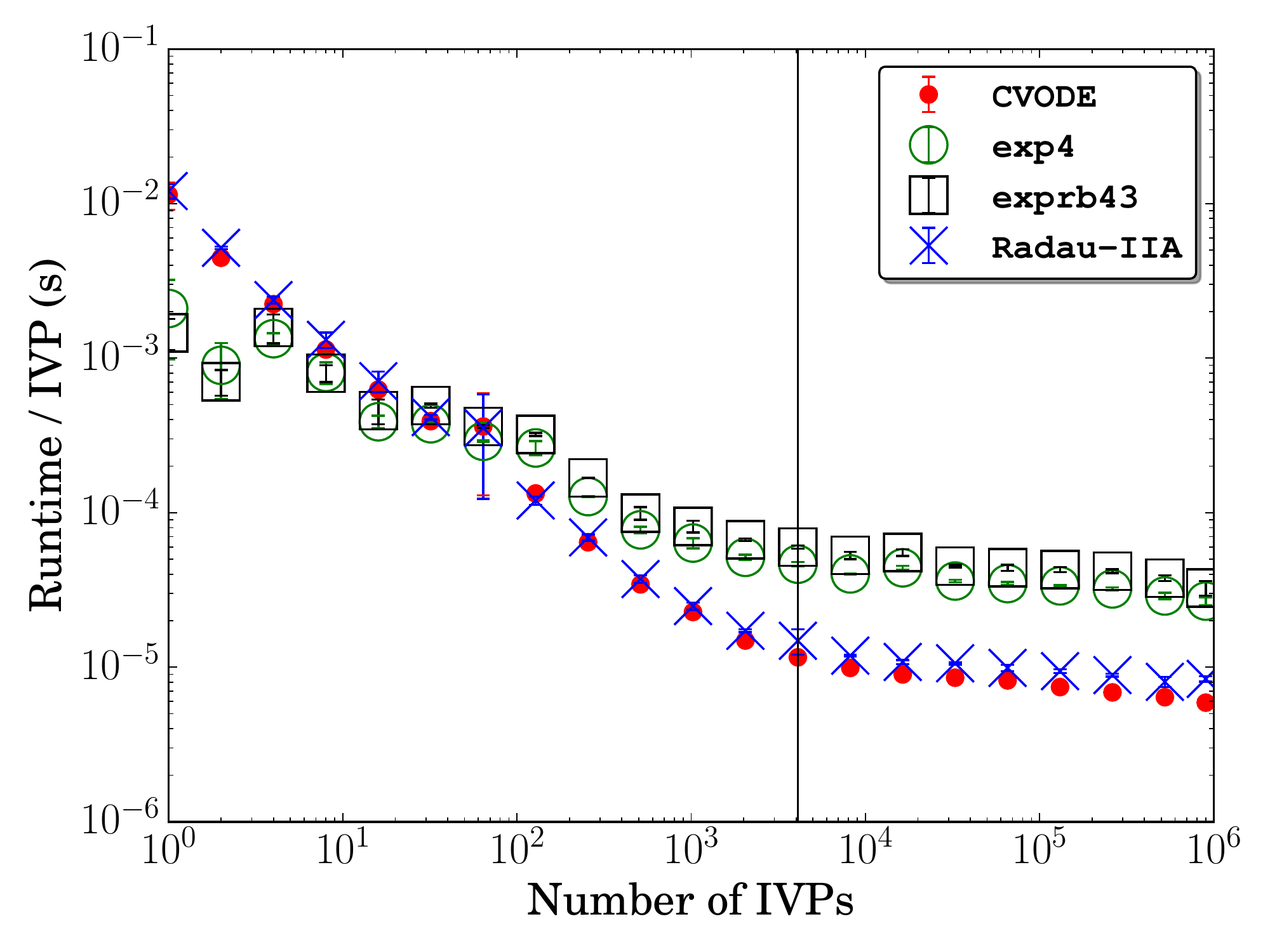}
      \caption{CPU performance results for $\Delta t = \SI{e-4}{\second}$}
      \label{F:h2_cpu_perf_large}
  \end{subfigure}\\
  \begin{subfigure}{0.49\textwidth}
      \includegraphics[width=\linewidth]{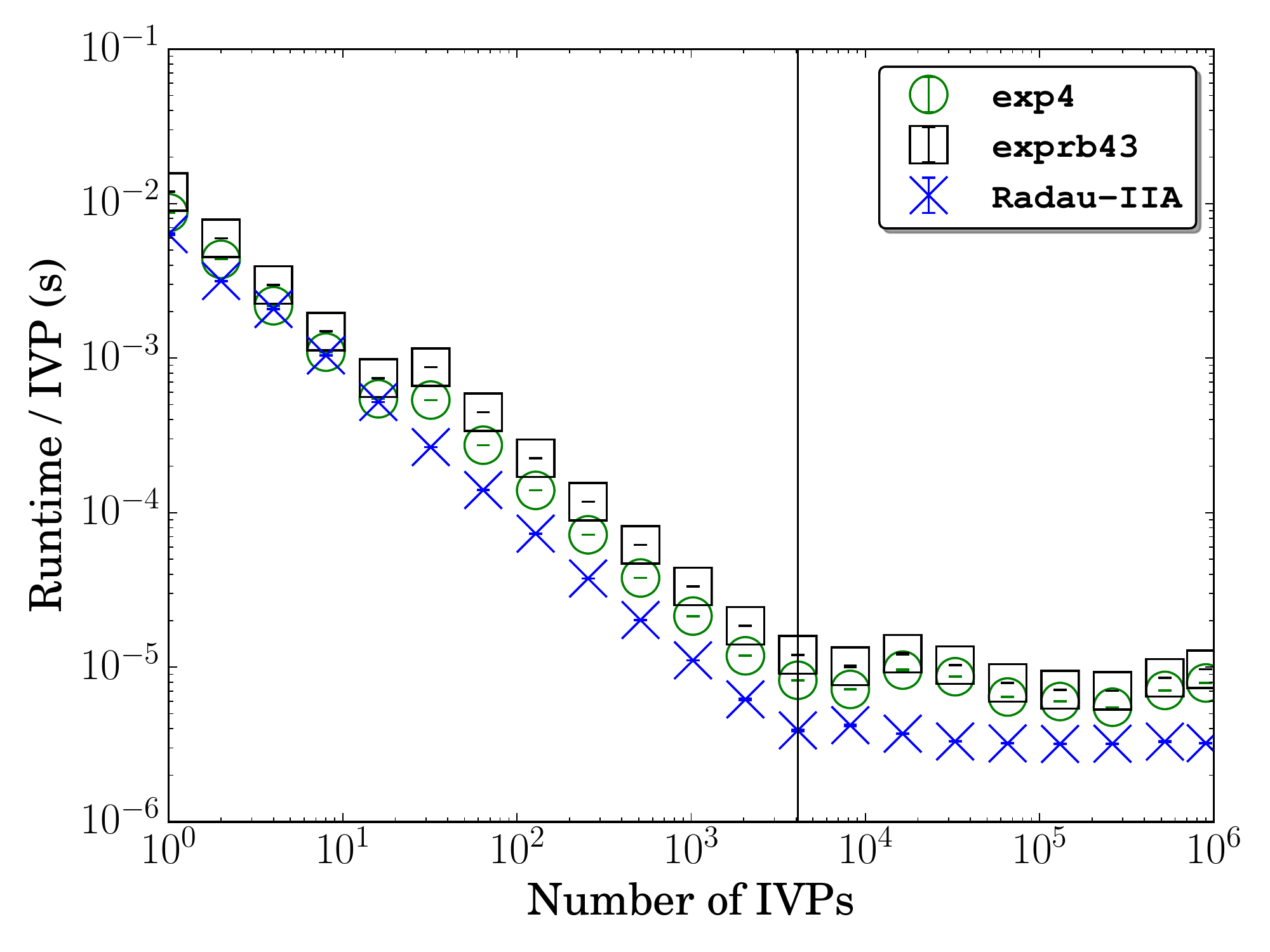}
      \caption{GPU performance results for $\Delta t = \SI{e-6}{\second}$}
      \label{F:h2_gpu_perf_small}
  \end{subfigure}
  \begin{subfigure}{0.49\textwidth}
      \includegraphics[width=\linewidth]{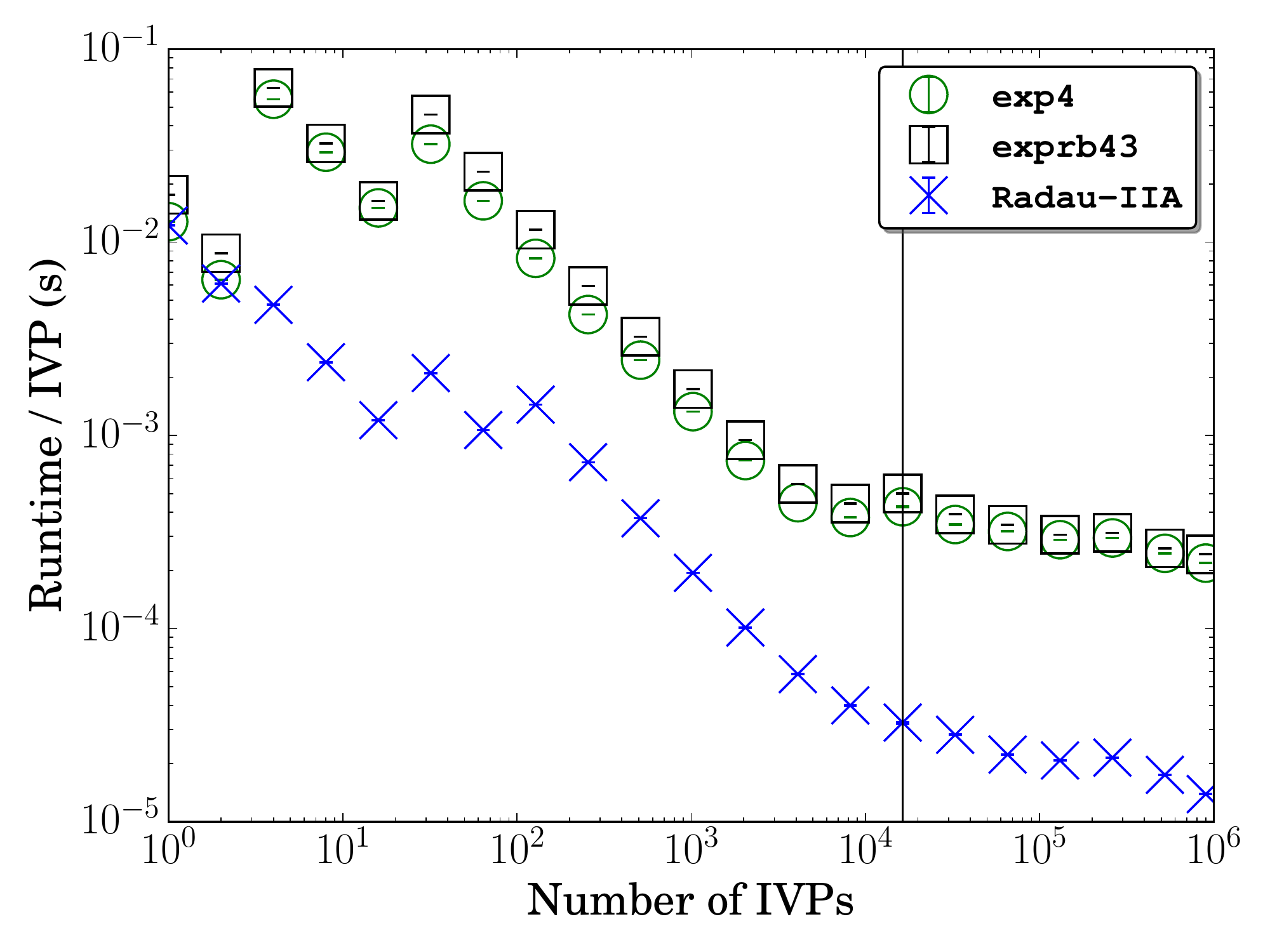}
      \caption{GPU performance results for $\Delta t = \SI{e-4}{\second}$}
      \label{F:h2_gpu_perf_large}
  \end{subfigure}
  \caption{Average runtimes of the integrators on the CPU and GPU, scaled by the number of IVPs, for the hydrogen model at two different global time-step sizes.
  Estimation of where the runtime per IVP levels off to a constant value (based on the results for \texttt{CVODE}\slash\texttt{Radau-IIA} for the CPU\slash GPU, respectively) is marked with a vertical line for all cases.
  Error bars indicate standard deviation.
  Data, plotting scripts, and figure files are available under CC-BY~\cite{paperscript:2017}.}
  \label{F:H2_perf}
\end{figure}

Figure~\ref{F:H2_perf} shows the runtimes of the CPU and GPU integrators for the hydrogen model.
In Fig.~\ref{F:h2_cpu_perf_small} the runtimes per IVP for the CPU integrators for a single global time-step of $\Delta t= \SI{e-6}{\second}$ decrease approximately linearly with the number of IVPs for small numbers of initial conditions (shown here on a log-log plot).
For small numbers of IVPs, the exponential integrators are faster than the implicit integration techniques due to the modest stiffness of the hydrogen model; even with many near-equilibrium states removed from the beginning of the PaSR database, the model is not particularly stiff for this small time-step size.
Larger numbers of IVPs begin to saturate the CPU resources, and the runtime per IVP levels off to a more constant value; vertical lines are shown in Fig.~\ref{F:H2_perf} where the relative change in runtime per IVP between successive data-points is first smaller than \SI{15}{\percent} (based on the results for \texttt{CVODE}\slash\texttt{Radau-IIA} for the CPU\slash GPU respectively).
Eventually, relatively more stiff conditions are encountered and the performance of the implicit integration techniques catches up and then surpasses that of the exponential integrators; \texttt{CVODE} is the most efficient solver on the CPU when solving more than \num{e4} IVPs; however, \texttt{CVODE} is only $\sim$\num{1.87}$\times$ faster than the slowest solver (\texttt{exprb43}) on the whole database.
Figure~\ref{F:h2_gpu_perf_small} shows the performances of the GPU integrators for the smaller global time-step size, which exhibit similar trends as the CPU solvers: a linearly decreasing solution cost that reaches a roughly constant value beyond \numrange{e3}{e4} IVPs.
Unlike for the CPU solvers, the GPU-based \texttt{Radau-IIA} performs faster than the exponential solvers for all numbers of IVPs.
As will be seen in Sec.~\ref{S:divergence}, both solver classes experience minimal thread divergence due to differing internal integration time-step sizes in this case.
Therefore, we conclude that the relatively slower runtimes per IVP for the exponential algorithms on the GPU results from thread divergence in the Arnoldi iteration---caused by varying Krylov subspace sizes between threads.

Figures~\ref{F:h2_cpu_perf_large} and \ref{F:h2_gpu_perf_large} show the performance of the integration algorithms on both platforms for the hydrogen model with a single larger global time step ($\Delta t=\SI{e-4}{\second}$).
The performances of the CPU integration algorithms show similar trends to those of the smaller global time-step size case: decreasing cost per IVP before reaching a more constant performance for higher numbers of IVPs.
The larger global time-step size induces additional stiffness, and the implicit solvers are more efficient for most numbers of IVPs; \texttt{CVODE} is again the most efficient CPU solver.
Figure~\ref{F:h2_gpu_perf_large} shows the performance of the GPU solvers for the larger global time-step size.
The exponential solvers exhibit significant spikes in computational cost when changing from 2--4 and 16--32 IVPs, with the latter mimicked somewhat by the implicit \texttt{Radau-IIA} solver.
A jump in solution cost between 2--4 IVPs is also present for the CPU exponential integrators, indicating stiffness as the primary cause.
On the other hand, between 16--32 IVPs the CPU exponential solvers exhibit only a very minor performance decrease, while the GPU-based \texttt{Radau-IIA} also shows a decrease in performance at the same point---a trend completely absent in the CPU \texttt{Radau-IIA} version.
These factors indicate that thread divergence also plays a key role in the performance trend here, and will be investigated further in Sec.~\ref{S:divergence}.
As in case of the smaller global time-step size, the \texttt{Radua-IIA} solver is the most efficient GPU algorithm in all cases.

\begin{figure}[htbp]
  \centering
  \begin{subfigure}{0.49\textwidth}
      \includegraphics[width=\linewidth]{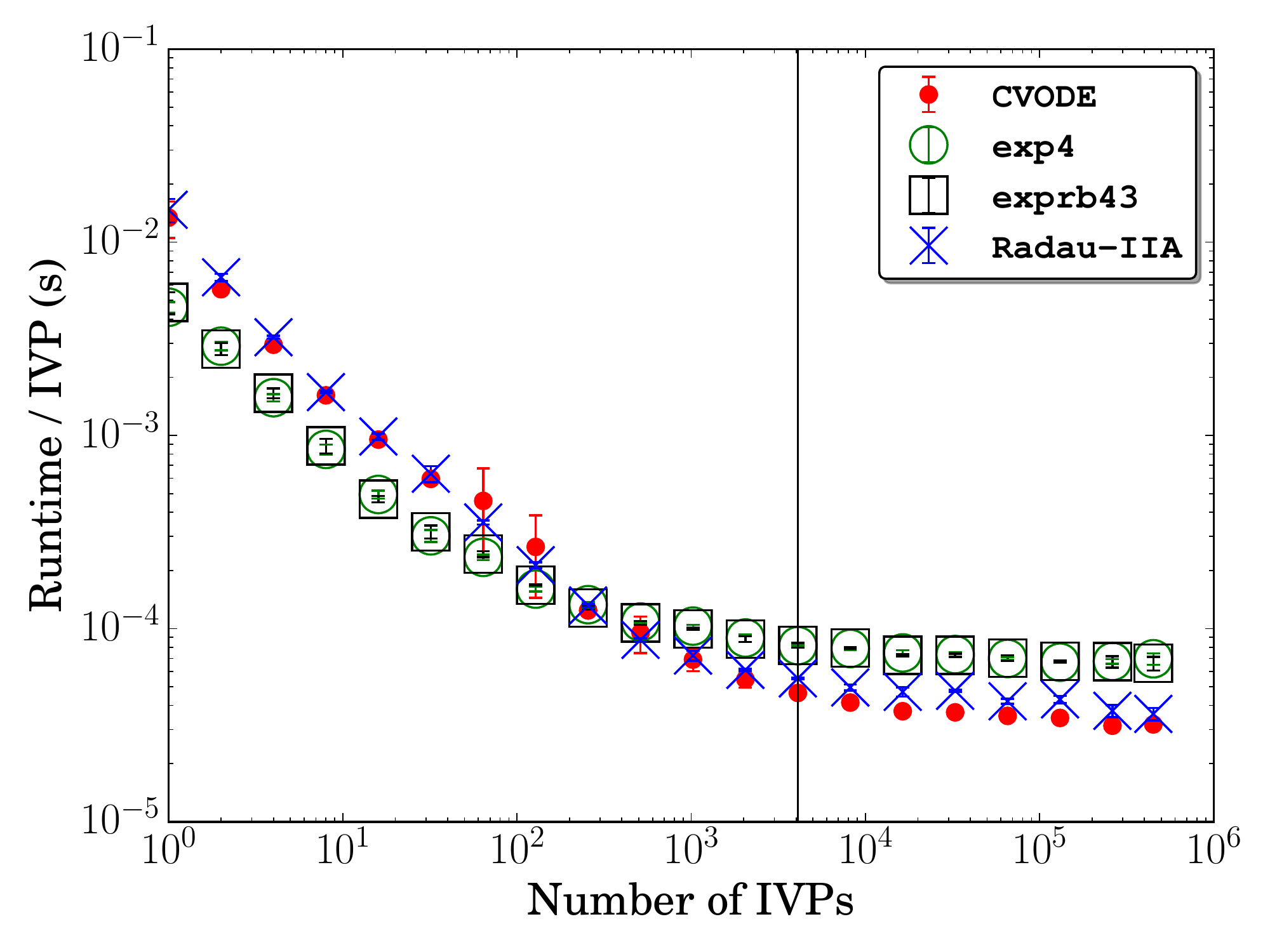}
      \caption{CPU performance results for $\Delta t = \SI{e-6}{\second}$}
      \label{F:ch4_cpu_perf_small}
  \end{subfigure}
  \begin{subfigure}{0.49\textwidth}
      \includegraphics[width=\linewidth]{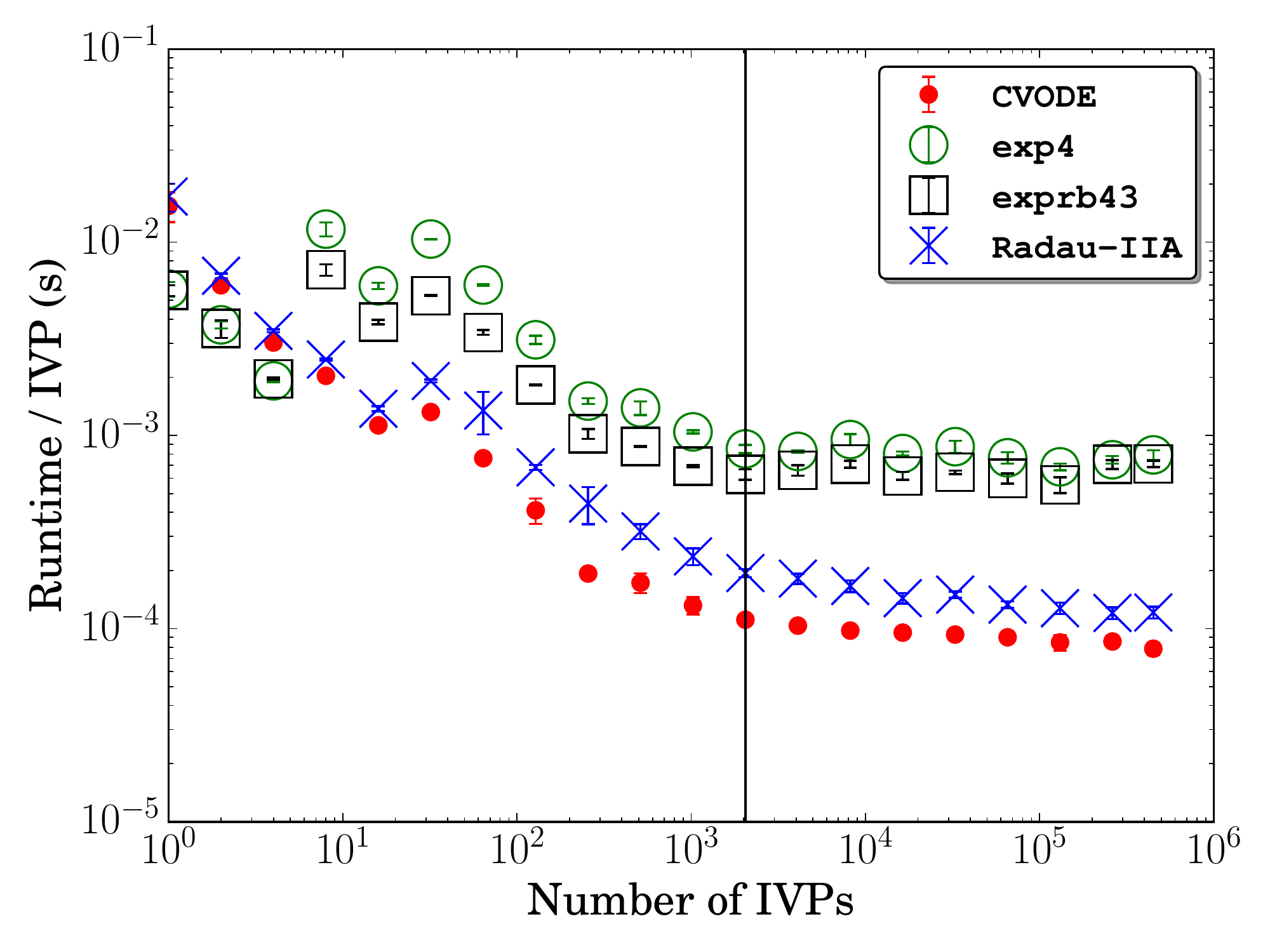}
      \caption{CPU performance results for $\Delta t = \SI{e-4}{\second}$}
      \label{F:ch4_cpu_perf_large}
  \end{subfigure}\\
  \begin{subfigure}{0.49\textwidth}
      \includegraphics[width=\linewidth]{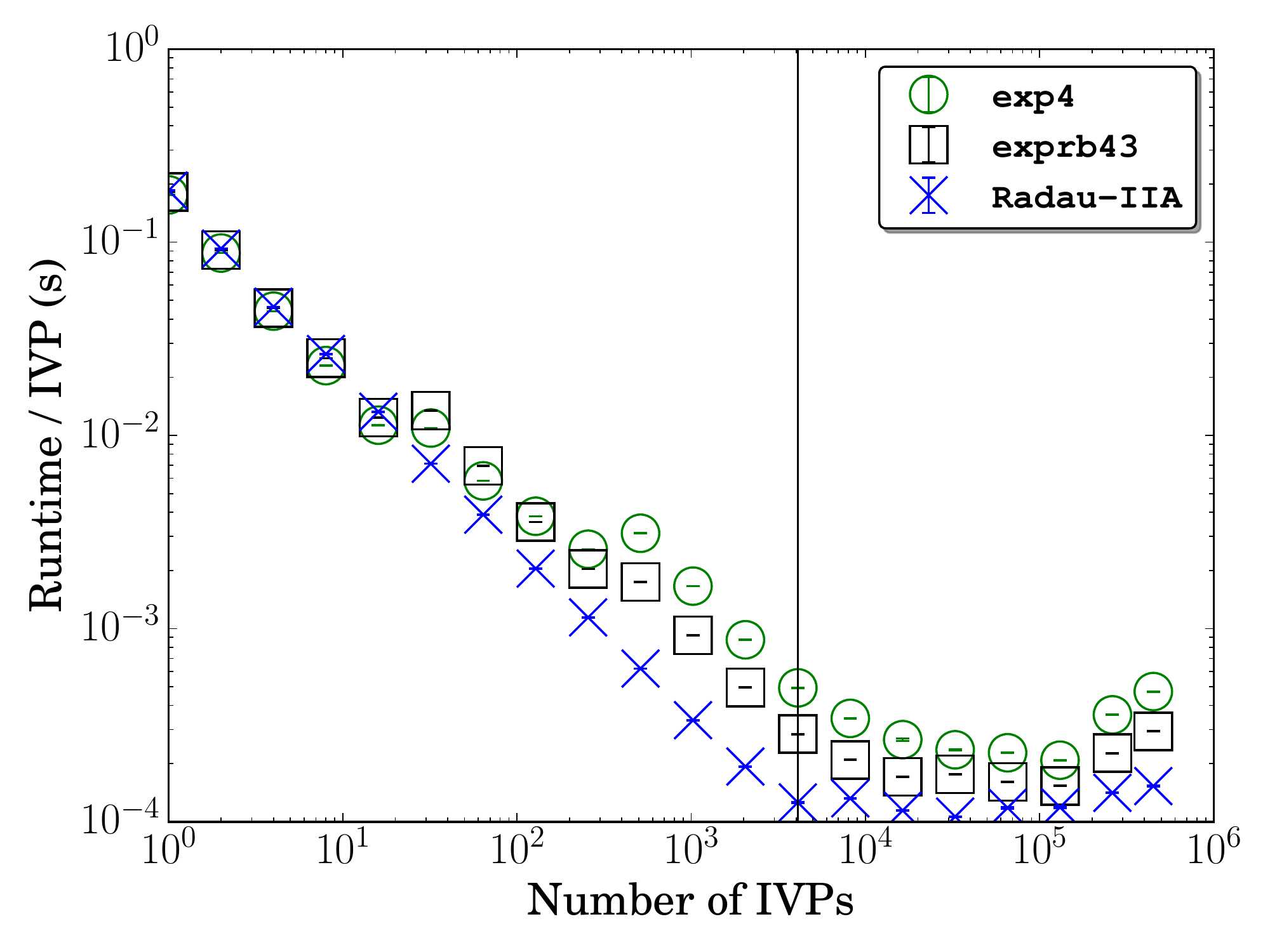}
      \caption{GPU performance results for $\Delta t = \SI{e-6}{\second}$}
      \label{F:ch4_gpu_perf_small}
  \end{subfigure}
  \begin{subfigure}{0.49\textwidth}
      \includegraphics[width=\linewidth]{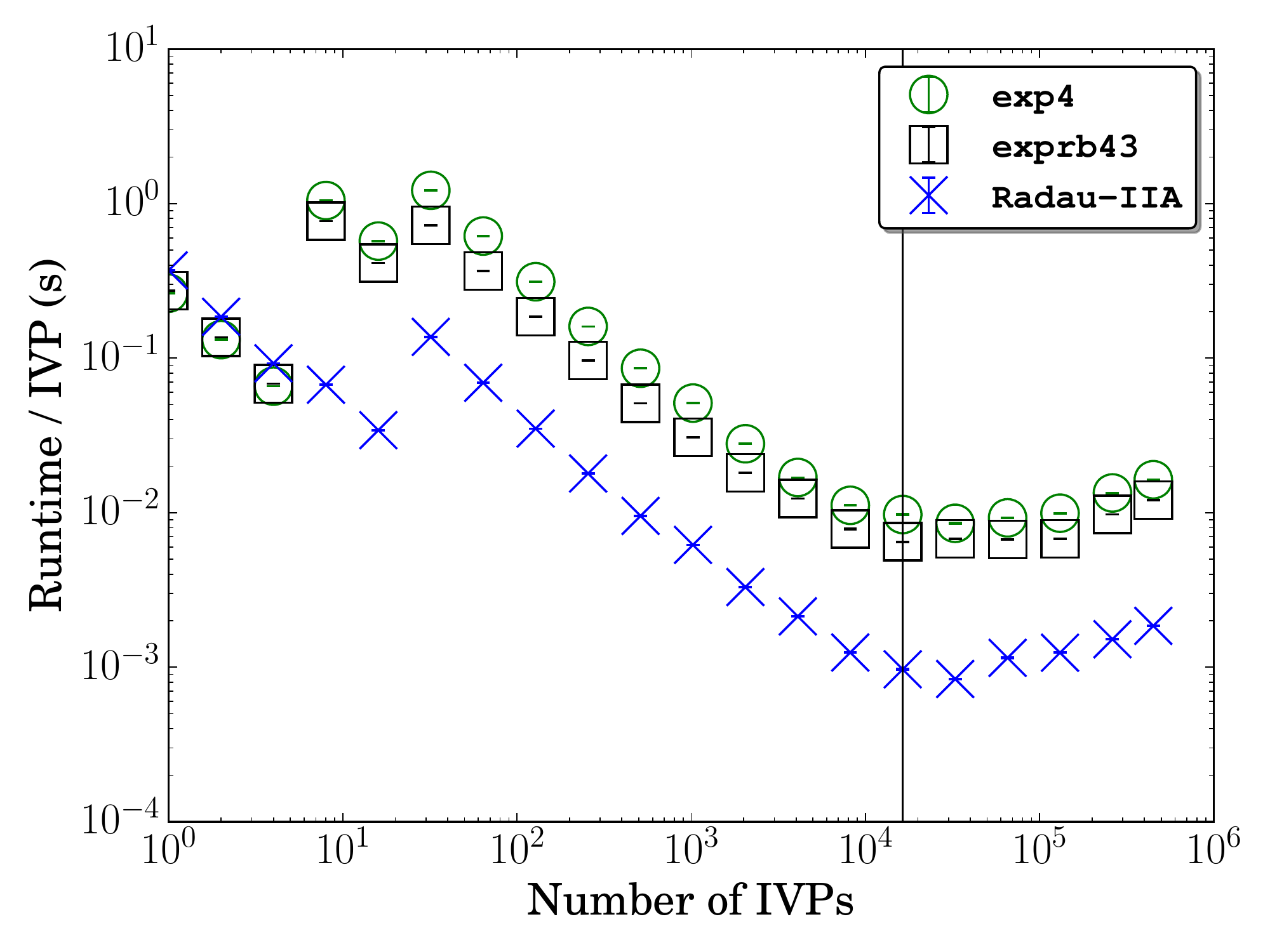}
      \caption{GPU performance results for $\Delta t = \SI{e-4}{\second}$}
      \label{F:ch4_gpu_perf_large}
  \end{subfigure}
  \caption{Average runtimes of the integrators, scaled by number of IVPs, on the CPU and GPU for the GRI-Mech 3.0 model at two different global time-step sizes.
  Estimation of where the runtime per IVP levels off to a constant value (based on the results for \texttt{CVODE}\slash\texttt{Radau-IIA} for the CPU\slash GPU respectively) is marked with a vertical line for all cases.
  Error bars indicate standard deviation.
  Data, plotting scripts, and figure files are available under CC-BY~\cite{paperscript:2017}.}
  \label{F:CH4_perf}
\end{figure}

Figure~\ref{F:CH4_perf} shows the runtime of the integrators for the GRI-Mech 3.0 model.
Similar to the hydrogen case for the smaller global time-step size, the CPU exponential integrators are more efficient (Fig.~\ref{F:ch4_cpu_perf_small}) for the near-equilibrium conditions at the beginning of the database.
For larger numbers of conditions, the implicit integrators are more efficient, and \texttt{CVODE} again performs the fastest.
Compared with the hydrogen model (Fig.~\ref{F:h2_cpu_perf_small}), the \texttt{CVODE} performs better than the exponential algorithms for the GRI-Mech 3.0 model with the small global time-step size (Fig.~\ref{F:ch4_cpu_perf_small}), reaching a speedup of \SI{2.18}{$\times$} over \texttt{exp4} on the whole database; this results from the higher stiffness present in the model.
This performance gap between the CPU implicit\slash exponential integrators increases for the larger global time-step size (Fig.~\ref{F:ch4_cpu_perf_large}); \texttt{CVODE} is \SI{10.1}{$\times$} faster than \texttt{exp4} on the whole database.
Comparing the performance of the CPU implicit solvers between the two kinetic models shows roughly an order-of-magnitude performance decrease for both global time-step sizes.
This phenomena, due largely to the increase in model size, is also seen for the \texttt{Radau-IIA} GPU solver for the smaller global time-step size; the performance of which decreases by just over an order of magnitude.
However, for the larger global time-step size, the GPU-based \texttt{Radau-IIA} solver performs roughly two orders-of-magnitude slower compared with the hydrogen case.
As will be examined in Sec.~\ref{S:divergence}, this dramatic decrease likely results from increased thread divergence in the \texttt{Radau-IIA} solver, as well as the increased memory traffic inherent in the larger model.

Unlike for the hydrogen model, the \texttt{exprb43} solver outperforms \texttt{exp4} with the GRI-Mech 3.0 model in almost all cases for the larger global time-step size for both the CPU and GPU.
Although the \texttt{exprb43} and \texttt{exp4} algorithms each require three exponential matrix function approximations per step, a single internal time step of \texttt{exprb43} is more expensive due to the extra chemical source term evaluations, matrix multiplications, and higher-order exponential matrix function requirement.
As such, the relatively simpler CPU \texttt{exp4} integrator outperforms the CPU \texttt{exprb43} integrator for the hydrogen model where there is relatively less stiffness.
However, as previously discussed the \texttt{exp4} algorithm may experience order reduction for stiff problems, and the \texttt{exprb43} algorithm typically outperforms \texttt{exp4} on both the CPU and GPU in the larger global time-step GRI-Mech 3.0 case as a result.


\subsection{CPU\slash GPU performance comparison}

Comparing the performance of CPU- and GPU-based integrators in a meaningful way is challenging.
First, the vastly different nature of the processing cores in each platform eliminates the possibility of comparing performance normalized by core count.
In addition, the floating-point operation count is not readily available for chemical kinetics integration---unlike many GPU-accelerated applications where the number of operations required to solve the problem is known, e.g., as in linear-algebra operations or fast Fourier transforms---which precludes comparing performance on the basis of floating-point operations per second (FLOPS).
Although the runtimes of the GPU integration algorithms can be directly compared with that of the CPU-based solvers (and often are), these figures do not provide much useful information.
For instance, if a GPU algorithm performs $\SI{10}{\times}$ faster than its equivalent on two six-core CPUs, how does this compare to two eight-core CPUs, etc.?

For researchers in numerical combustion, two issues stand out as particularly important for performance evaluation: runtime and cost.
As established in Sec.~\ref{sec:Intro}, large-scale reactive-flow simulations with realistic chemical kinetic models are extremely computationally expensive, and remain outside of the capabilities of most in the field.
With this in mind, we ask, for a given simulation, what is the effect on the overall runtime of adding more CPU cores compared with adding GPU accelerators?
In addition, if a budget is allocated to expand available computational resources, how might these funds be best allocated?
To answer these questions, we derived an estimate of the number of CPU cores required for equivalent performance on the GPU.

A nominal performance metric for both the CPU- and GPU-based integration algorithms must first be obtained.
As the most efficient solvers in all cases with large numbers of IVPs are \texttt{CVODE} for the CPU and \texttt{Radau-IIA} for the GPU, these algorithms will be considered the performance benchmarks.
Furthermore, most large-scale simulations consist of millions of cells (or more), and therefore we only consider the performance limit of each algorithm (i.e., the cost per IVP of each algorithm in the region where this cost reaches an approximately constant value).
To this end, the previously discussed threshold---the first relative change in runtime per IVP between successive data-points smaller than \SI{15}{\percent} (based on \texttt{CVODE}\slash\texttt{Radau-IIA} for the CPU\slash GPU accordingly)---is used, and marked as vertical lines on Figs.~\ref{F:H2_perf} and \ref{F:CH4_perf}.
The cost per IVP above and including these thresholds was averaged and forms our nominal performance measure.
The CPU performance measure must also be normalized by the total number of cores used: \num{40}.
Table~\ref{T:cpu_equiv} presents the ratios of these performance measures, which give estimates for the number of CPU cores required to equal the GPU performance for the cases studied.
The GPU is roughly equivalent to \num{12} or more CPU cores for all cases except GRI-Mech 3.0 with the larger global time-step size, and equivalent to at most \num{38} cores for the hydrogen case with the smaller global time-step size.
With the increasing size of the chemical kinetic model, the equivalent CPU core count of the GPU \texttt{Radau-IIA} solver drops significantly.
As will be discussed in Sec.~\ref{S:divergence}, this drop in performance is primarily due to higher memory traffic requirements, however increased levels of thread divergence also play a role.
Although this work represents the current state-of-the-art for implicit integration of stiff chemical kinetic IVPs on the GPU, it is clear that more effort is required to improve GPU performance for larger chemical kinetic models.
Approaches to mitigate these issues will be discussed in the subsequent section.

\begin{table}[htbp]
\centering
\begin{tabular}{@{}l S[table-format=2.1] S[table-format=2.1] @{}}
\toprule
\multirow{2}{*}{Global time-step size} & \multicolumn{2}{c}{\# equivalent CPU cores} \\ \cmidrule{2-3}
 & {Hydrogen} & {GRI-Mech 3.0} \\
\midrule
\SI{e-6}{\second} & 38 & 12 \\
\SI{e-4}{\second} & 15 & 3 \\
\bottomrule
\end{tabular}
\caption{The number of CPU cores (roughly) required for equivalent performance to a single GPU for the combinations of chemical kinetic models and global time-step sizes studied.}
\label{T:cpu_equiv}
\end{table}

At the time of writing, the ten-core Intel Xeon E5-4640 v2 CPU used in this study was listed for a recommended customer price of \SI{2725}[\$]{}~\cite{intel_price}, while a new Tesla C2075 GPU is available for $\sim$\SI{1400}[\$]{}~\cite{gpu_price}.
These prices are only rough estimates of the actual cost of these devices, since the actual price for the Intel CPU may be significantly less in a configured server node, while the Tesla C2075 is no longer sold directly by NVIDIA---thus the prices are variable.
Furthermore, the performance decrease using an older, cheaper CPU (e.g., the Intel Xeon X5650 used as host processor for the GPU simulations in this work) may not be that large.
However, combined with the equivalent core counts in Table~\ref{T:cpu_equiv}, this information suggests that the Tesla C2075 is a reasonable investment to supplement computing power for chemical-kinetic integration in large-eddy simulations.

\subsection{Effects of thread divergence and memory traffic}
\label{S:divergence}

Thread divergence and memory traffic are two performance concerns particularly important for chemical kinetics integration on GPU and SIMT platforms.
Slowdown due to memory traffic for a GPU integration algorithm implemented on a per-thread basis primarily results from the small amount of on-chip memory available.
Implicit integration algorithms, which typically require storage of the Jacobian matrix and\slash or factorized forms thereof, can quickly overwhelm the registers and L1 cache memory available to each thread and cause many slow global memory accesses.
Reformulating the chemical kinetic equations to generate sparse Jacobian matrices~\cite{Schwer2002270} would greatly benefit GPU-based integration algorithms due to the reduced memory requirements, and in addition enable use of sparse multiplication\slash factorization algorithms (from which a CPU-based algorithm would also benefit); this is a planned improvement to the \texttt{pyJac} software~\cite{Niemeyer:2016aa,niemeyer_2016_51139}.\footnote{Bisetti~\cite{Bisetti:2012jw} demonstrated a method to exploit the underlying sparsity of a dense mass-fraction-based constant-pressure Jacobian matrix (used in this study) to accelerate Jacobian-vector multiplications; however, a reformulation is still more attractive as it enables sparse-LU factorization.}
Further, the Tesla C2075 GPU used in this study was originally released nearly five years ago and is several generations old; the newer Tesla K40 is available for a similar price, \SI{2950}[\$]{}~\cite{k40_price}, as the Xeon E5-4640 v2 CPU used in this study, and has $2\times$ registers available per block~\cite{NVIDIA:2015aa} and $6.4\times$ as many CUDA cores~\cite{k40_specs} as the Tesla C2075 used.
Using a newer GPU model could significantly improve solver performance for larger models by relieving the scarcity of on-chip memory in a per-thread approach.
Finally, a per-block approach may be required to efficiently integrate the largest models on the GPU, due to the much higher amount of cache memory allocated for each IVP solution.

The performance penalty due to thread divergence depends both on the cost of the divergent branches as well as the proportion of the warp that executes each branch.
For example, if only one thread in a warp executes an expensive branch (e.g., a Jacobian update), the rest of the warp remains idle during that time, and the SM may become severely underutilized.
To investigate the effects of thread divergence further, we adopted a modified version of the quantification of thread divergence of Niemeyer and Sung~\cite{Niemeyer:2014aa}:
\begin{equation}
	D = 1 - \frac{\sum_{i=1}^{32}{d_i}}{32 \times \max\limits_{i = 1, \dots, 32} d_i} \;,
	\label{eqn:divergence}
\end{equation}
where $d_i$ is the number of internal integrator time steps taken to reach the global time step by thread $i$ in a warp (which consists of 32 threads).
$D$ represents the similarity of internal time step counts across threads in a warp---a significant source of thread divergence.
If all threads in a warp use identical internal integration time steps and thus the warp experiences no thread divergence from this source, then $D = 0$; however, if a warp experiences an unbalanced number of internal integration time steps, then $D \to 1$.
Differing internal time-step sizes are not the only source of thread divergence for the GPU integration algorithms.
For instance, threads in a warp may use different Krylov subspace sizes for the exponential integrators or different numbers of Newton iterations for the \texttt{Radau-IIA} solver.
Indeed, Sec.~\ref{S:perf} notes that we suspect thread divergence from differing Krylov subspace sizes as the reason the exponential solvers are less efficient for small numbers of IVPs for the hydrogen model with the small global time-step size.
However, these operations clearly cost less than an entire internal integration step (in which they are embedded) and thus we look only at the thread divergence of internal integration time steps.
Thread divergence of such operations within an internal integration step could play an important role and will be investigated in our future work.

\begin{figure}[htbp]
  \centering
  \begin{subfigure}{0.49\textwidth}
      \includegraphics[width=\linewidth]{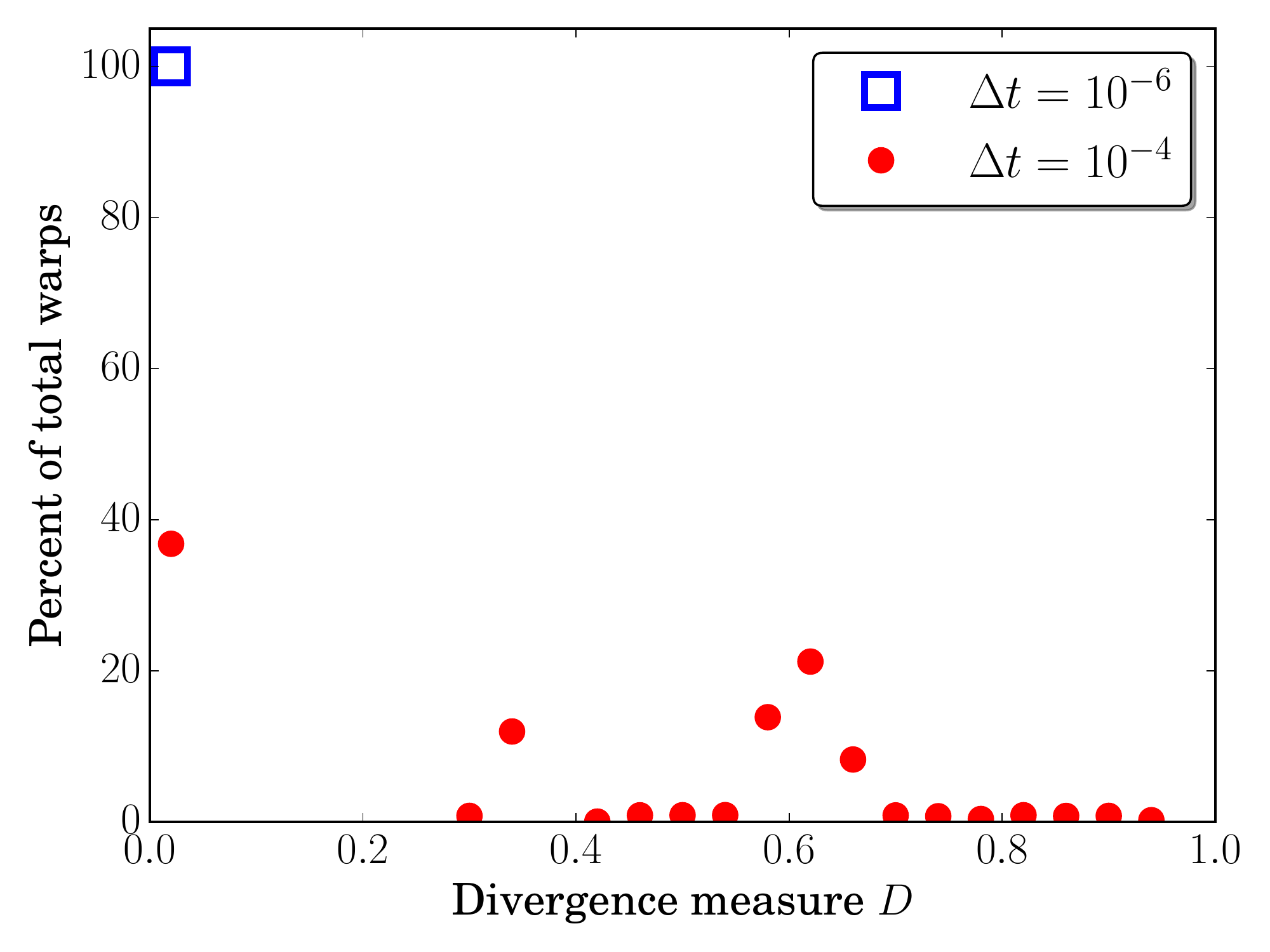}
      \caption{\texttt{Radau-IIA} solver for hydrogen model}
      \label{F:Rad_div_h2}
  \end{subfigure}
  \begin{subfigure}{0.49\textwidth}
      \includegraphics[width=\linewidth]{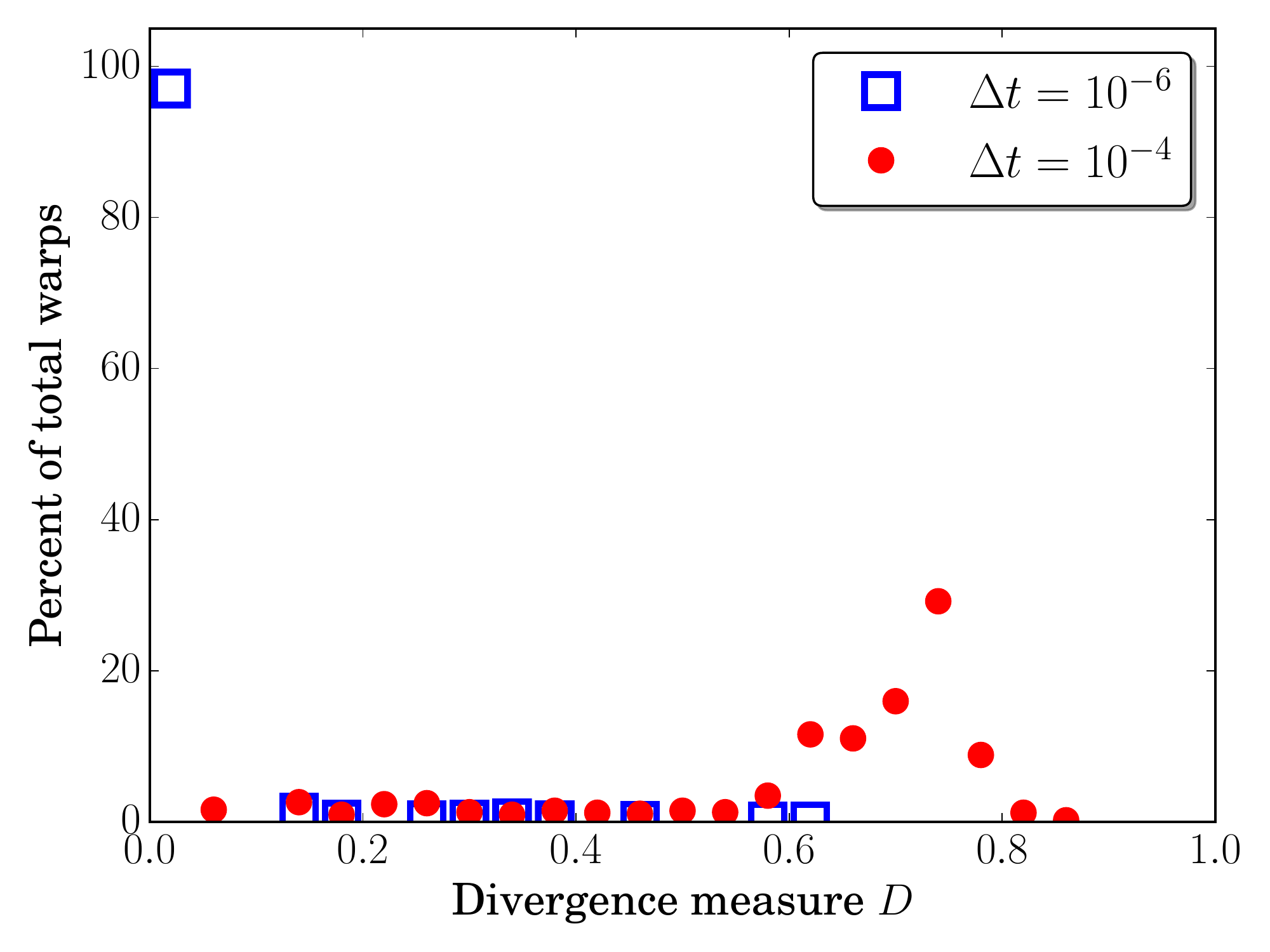}
      \caption{\texttt{Radau-IIA} solver for GRI-Mech 3.0 model}
      \label{F:Rad_div_gri}
  \end{subfigure}
  \\
  \begin{subfigure}{0.49\textwidth}
      \includegraphics[width=\linewidth]{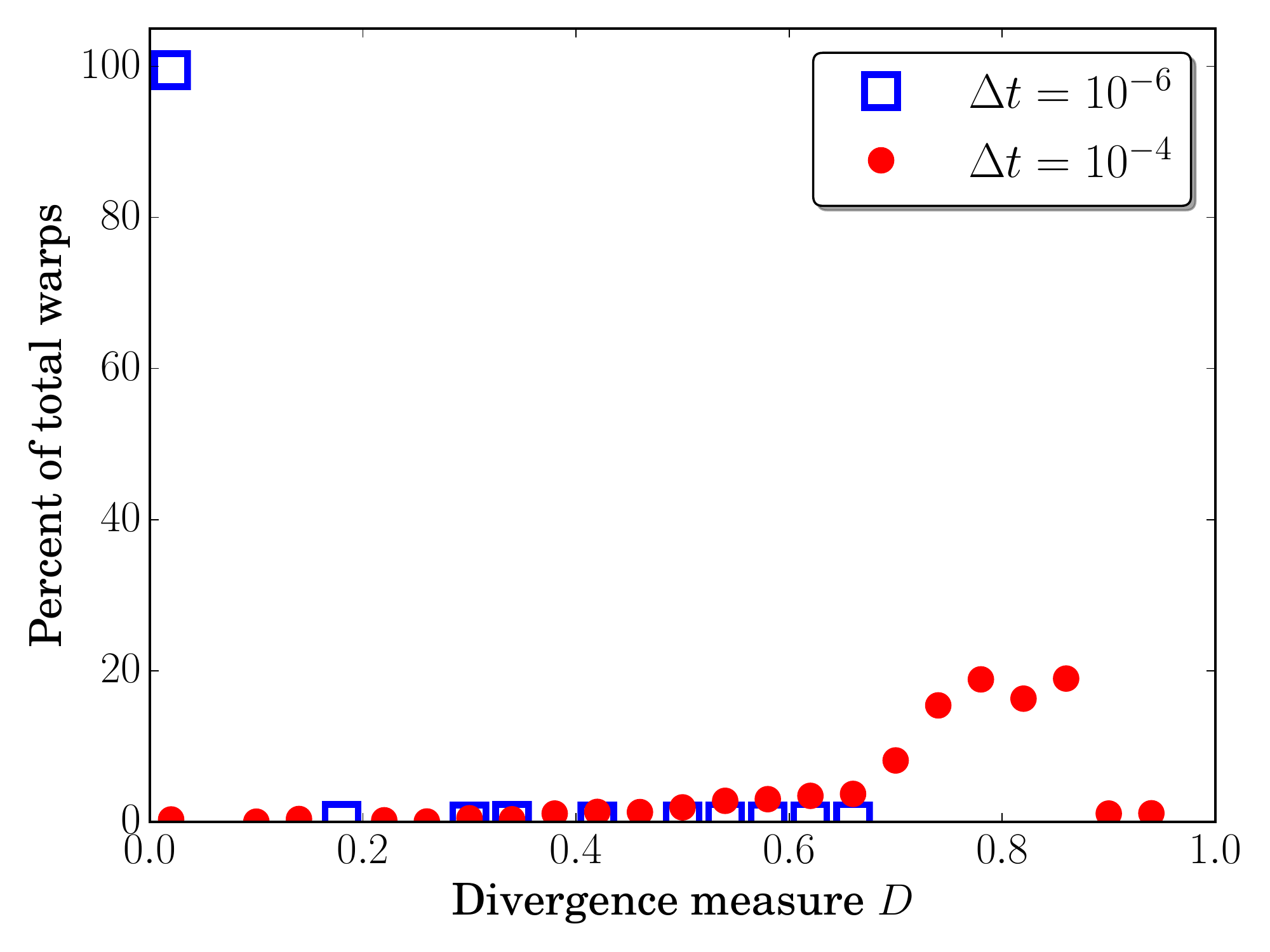}
      \caption{\texttt{exprb43} solver for hydrogen model}
      \label{F:exprb43_div_h2}
  \end{subfigure}
  \begin{subfigure}{0.49\textwidth}
      \includegraphics[width=\linewidth]{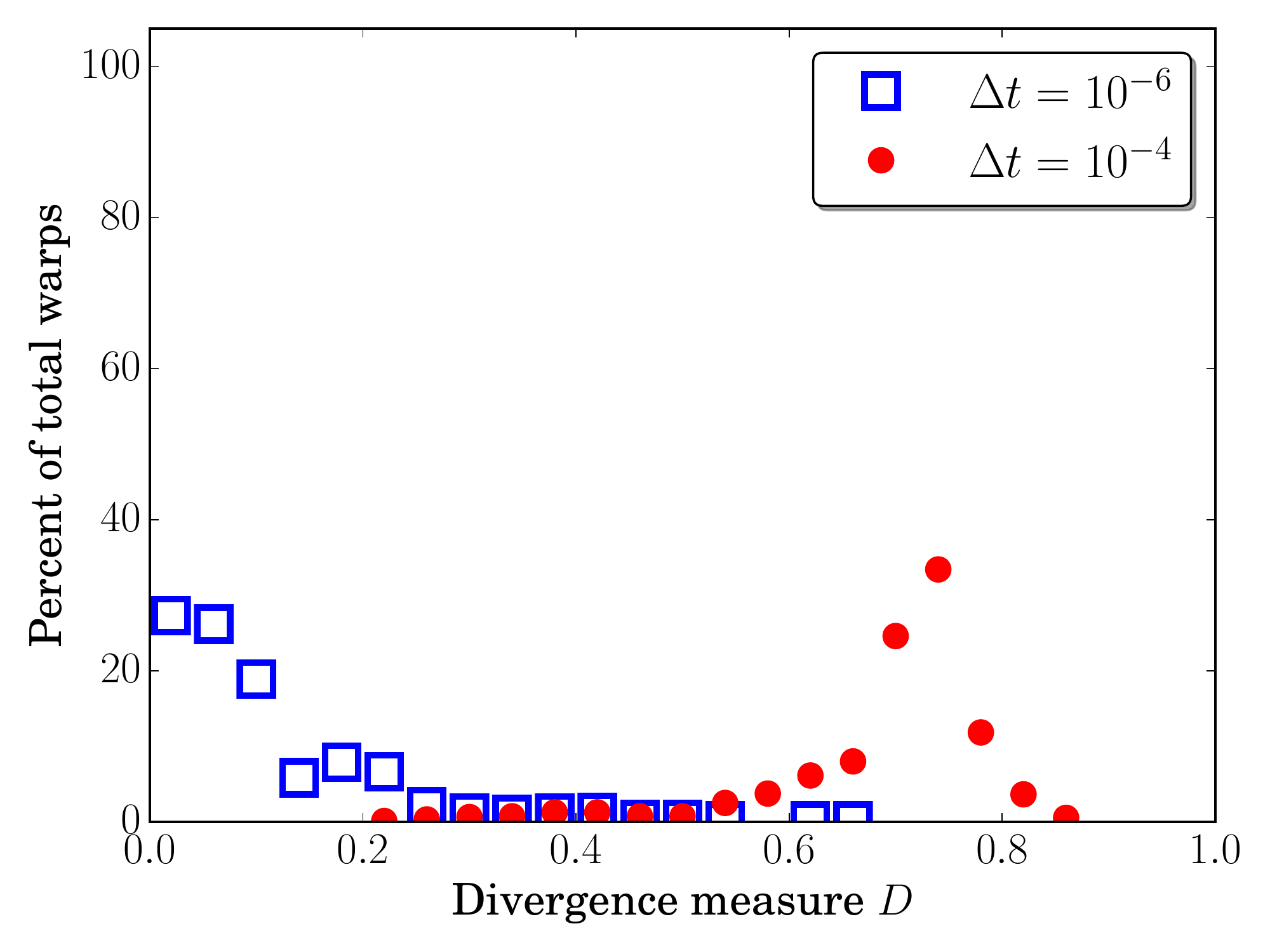}
      \caption{\texttt{exprb43} solver GRI-Mech 3.0 model}
      \label{F:exprb43_div_ch4}
  \end{subfigure}
  \\
  \begin{subfigure}{0.49\textwidth}
      \includegraphics[width=\linewidth]{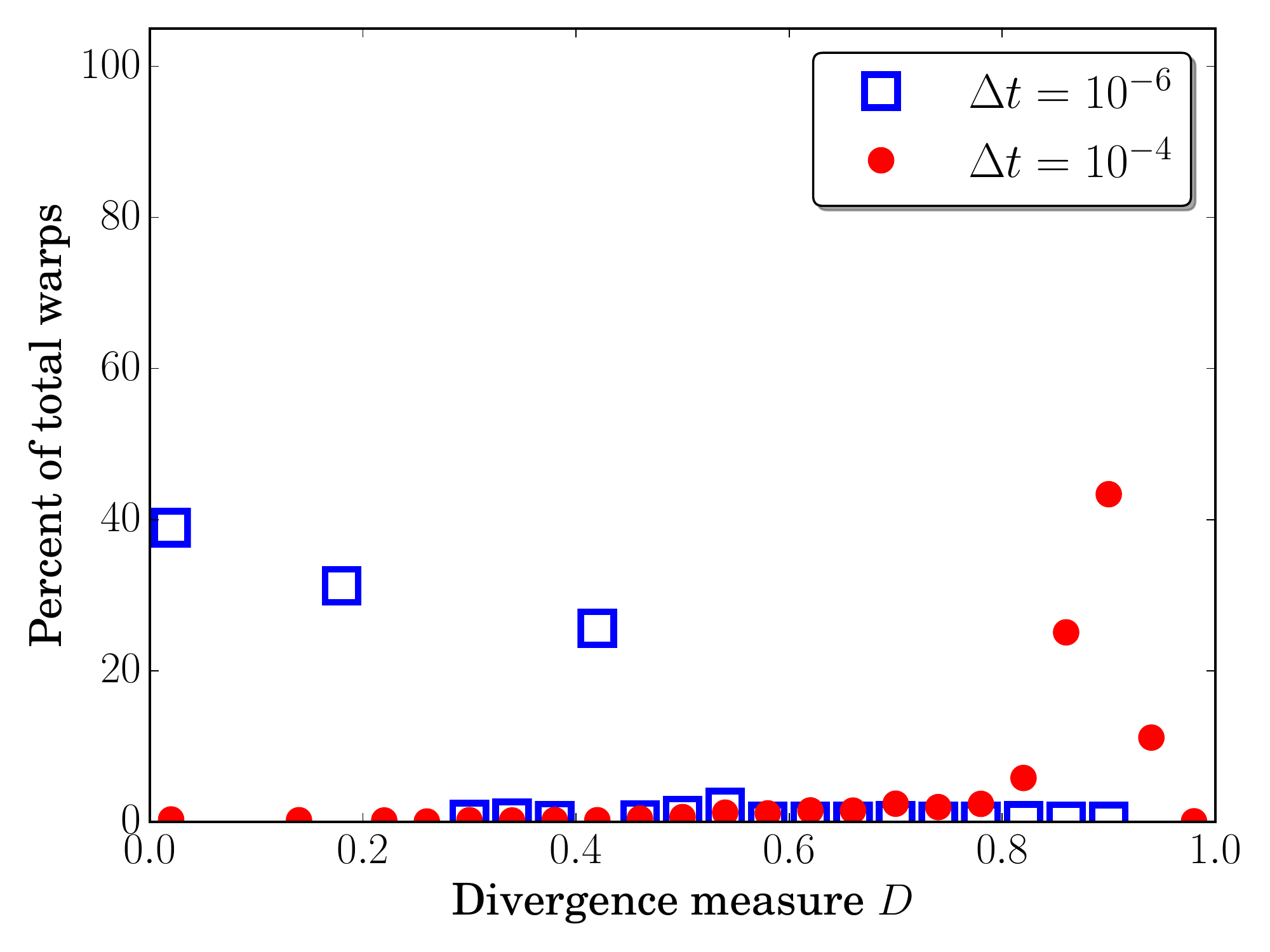}
      \caption{\texttt{exp4} solver hydrogen model}
      \label{F:exp4_div_h2}
  \end{subfigure}
  \begin{subfigure}{0.49\textwidth}
      \includegraphics[width=\linewidth]{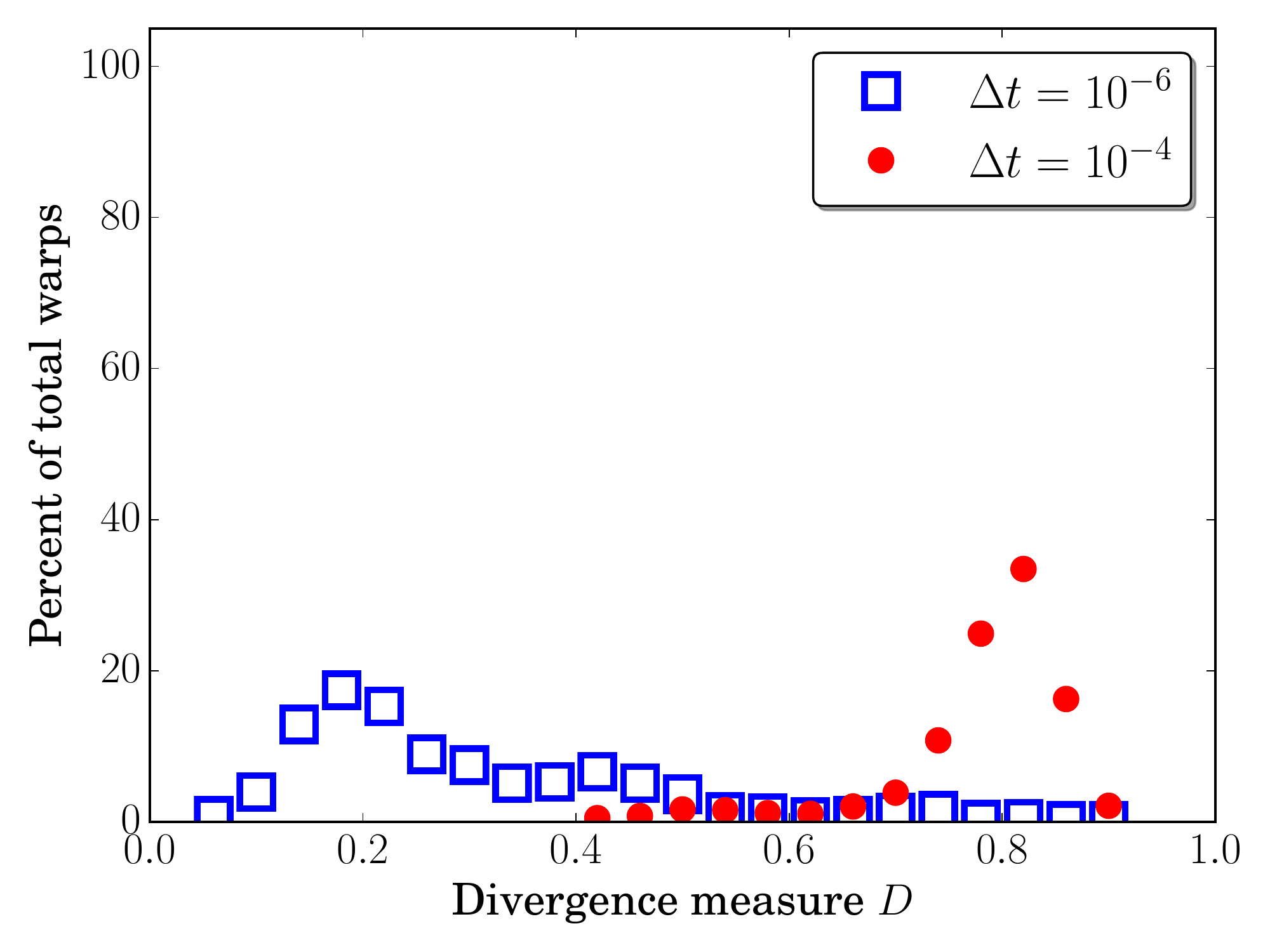}
      \caption{\texttt{exp4} solver GRI-Mech 3.0 model}
      \label{F:exp4_div_ch4}
  \end{subfigure}
  \caption{Thread divergence estimate for the three solvers for both models and global time-step sizes.
  Data, plotting scripts, and figure files are available under CC-BY~\cite{paperscript:2017}.}
  \label{F:divergence}
\end{figure}

Figures~\ref{F:Rad_div_h2} and~\ref{F:Rad_div_gri} show the distribution of the divergence measure $D$ for the \texttt{Radau-IIA} solver with both global time-step sizes and kinetic models when run on \num{262144} IVPs, spread across \num{8192} warps.
For both kinetic models with the smaller global time-step size, nearly \SI{100}{\percent} of the warps had a divergence measure near zero.
Increasing the global time-step size causes the number of warps with high levels of thread divergence (e.g. $D > 0.5$) to increase for both models.
For the hydrogen model, over \SI{40}{\percent} of warps were between $D=0.55$ and $D=0.65$, and the approximate equivalent CPU core-count (Table~\ref{T:cpu_equiv}) dropped by \SI{2.5}{$\times$} between the small and large global time-step sizes.
Further, over \SI{75}{\percent} of warps were between $D=0.6$ and $D=0.8$ for the GRI-Mech 3.0 model for the larger global time-step size, and subsequently a higher drop in performance of \SI{4}{$\times$} occurred.
This observation motivates future work aimed at developing strategies to reduce thread divergence.
Potential solutions include adopting an IVP per-block approach~\cite{Stone:2013aa}, reordering IVPs to increase similarity of stiffness inside a warp, or synchronizing internal time-step sizes between threads in a warp.
However, Figs.~\ref{F:Rad_div_h2} and \ref{F:Rad_div_gri} do not explain the drop in equivalent core count between the hydrogen model and the GRI-Mech 3.0 model for the smaller global time-step size.
The minimal thread divergence of the \texttt{Radau-IIA} solver for both models at the smaller global time-step size suggests that this drop in performance is primarily caused by the increased memory traffic of the larger model, as well potential thread divergence inside the internal integration step; this further motivates development of a sparse version of the \texttt{pyJac}~\cite{niemeyer_2016_51139,Niemeyer:2016aa} software.

Figures~\ref{F:exprb43_div_h2} and \ref{F:exprb43_div_ch4} show the divergence levels of the \texttt{exprb43} GPU solver.
Similar to the \texttt{Radau-IIA} solver, nearly \SI{100}{\percent} of warps for the \texttt{exprb43} solver have no thread divergence due to differing internal integration step sizes for the hydrogen model.
The \texttt{exprb43} thread divergence levels increase somewhat for the GRI-Mech 3.0 model with the smaller time-step size; \SI{27}{\percent} of warps still had a divergence measure of $D=0$, but nearly \SI{63}{\percent} of the warps had divergence measures between $D=0.05$ and $D=0.2$.
With the larger time-step size, the \texttt{exprb43} solver experiences significantly more thread divergence for both models.
The divergence measure distribution is fairly similar to that of the \texttt{Radau-IIA} solver for the GRI-Mech 3.0 model, but most warps experience a divergence measure of $D \sim 0.8$ for the hydrogen model (versus $D \sim 0.6$ for the \texttt{Radau-IIA} solver).
The semi-implicit solvers deal with stiffness less efficiently, and end up using a greater range of internal time-step sizes between conditions of varying stiffness.
This results in an increase in thread divergence levels due to differing internal time-step sizes.

The relatively worse stiffness handling of the \texttt{exp4} method is also apparent in Figs.~\ref{F:exp4_div_h2} and \ref{F:exp4_div_ch4}; in most cases, significantly more thread divergence is seen for \texttt{exp4} than for either of the other two solvers.
The \texttt{exp4} algorithm is the only solver to show significant thread divergence even for the hydrogen model for the smaller global time-step size.
Further, the \texttt{exp4} algorithm experiences more thread divergence than the \texttt{exprb43} for both models at the larger global time-step size.

\subsection{Effect of using a finite-difference-based chemical kinetic Jacobian}

While it is well established that using an analytical Jacobian matrix can significantly accelerate chemical kinetics integration on the CPU~(e.g., \cite{Lu:2009gh,stone2014comparison,Schwer2002270}), relatively little study has been directed at use of a GPU-based analytical Jacobian.
Dijkmans et al.~\cite{Dijkmans:2014bb} used a GPU-based analytical Jacobian code to accelerate various CPU-based chemical kinetics integration schemes, and our own previous works~\cite{Niemeyer:2016aa,Niemeyer:2015ws} have detailed the performance of \texttt{pyJac}.
However, to our knowledge no work using an analytical Jacobian for GPU-based chemical kinetics integration has been published.
In this section, we explore the relative performance benefits of the analytical Jacobian compared with a first-order finite-difference Jacobian on both the CPU and GPU.
The exponential methods require an exact Jacobian matrix (rather than an approximation as given by finite-difference methods), so their performance was not considered in this section.

\begin{figure}[htbp]
  \centering
  \begin{subfigure}{0.49\textwidth}
      \includegraphics[width=\linewidth]{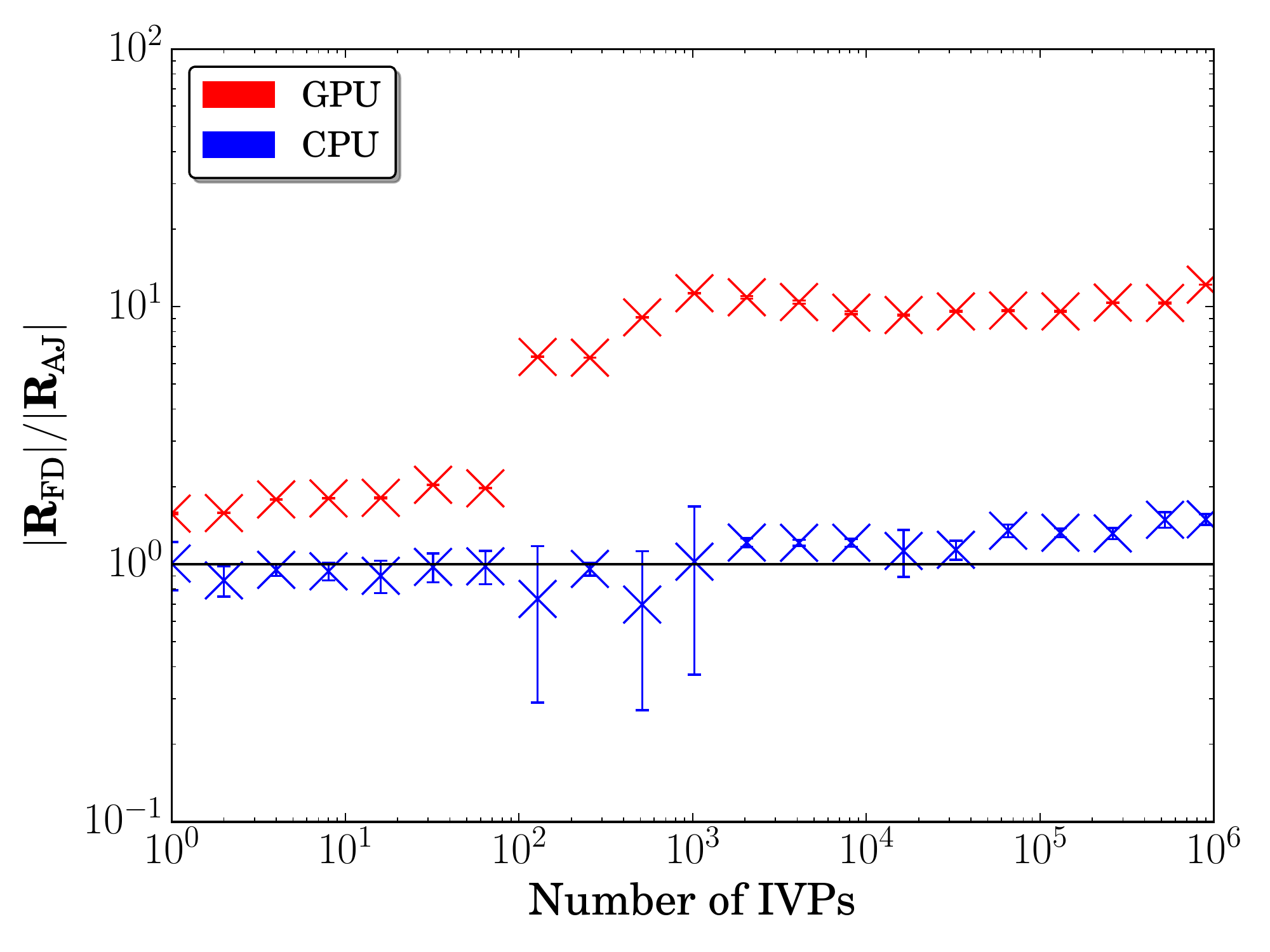}
      \caption{Hydrogen model with $\Delta t = \SI{1e-6}{\second}$}
      \label{F:AJ_h2_small}
  \end{subfigure}
  \begin{subfigure}{0.49\textwidth}
      \includegraphics[width=\linewidth]{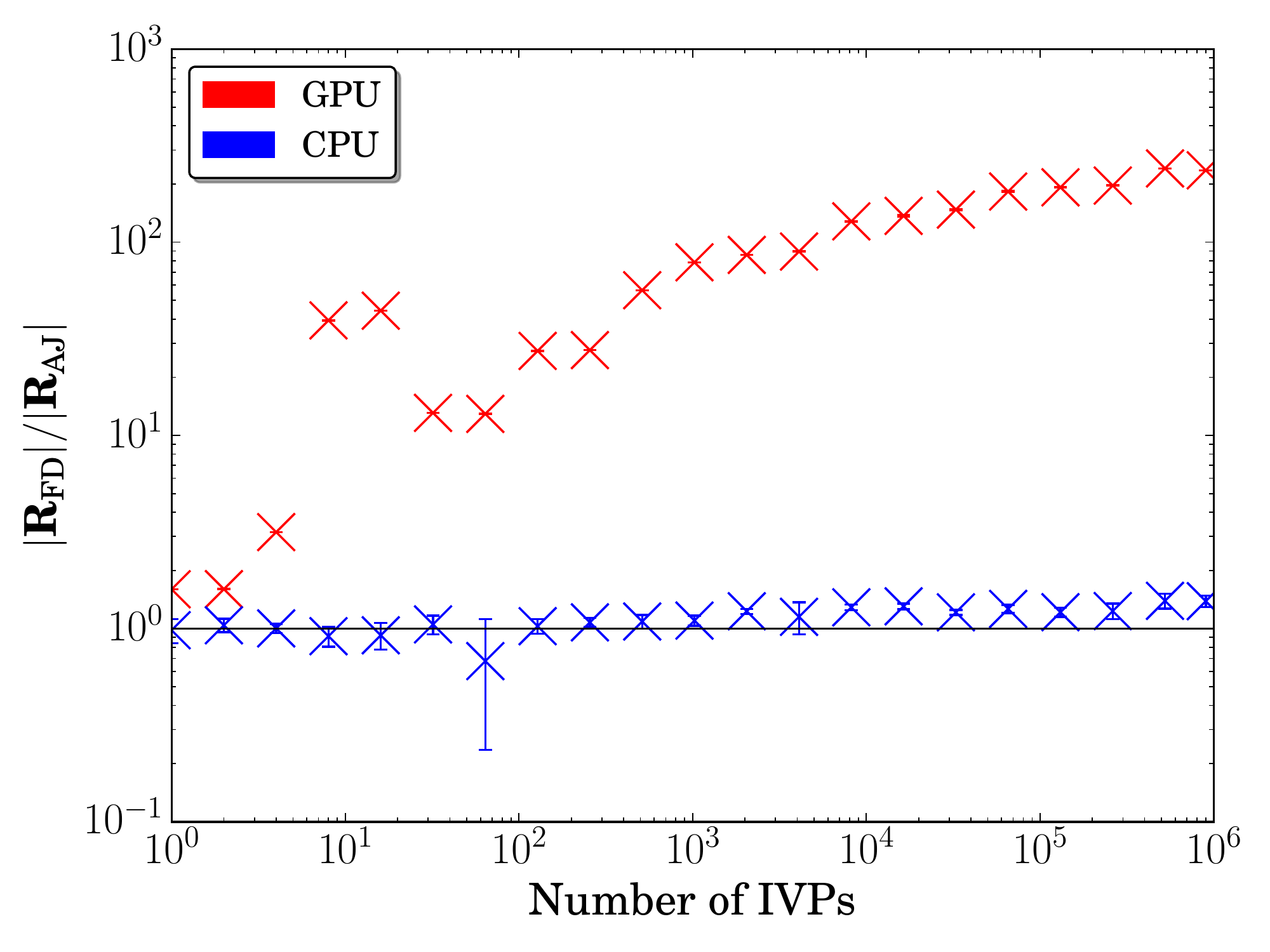}
      \caption{Hydrogen model with $\Delta t = \SI{1e-4}{\second}$}
      \label{F:AJ_h2_large}
  \end{subfigure}
  \\
  \begin{subfigure}{0.49\textwidth}
      \includegraphics[width=\linewidth]{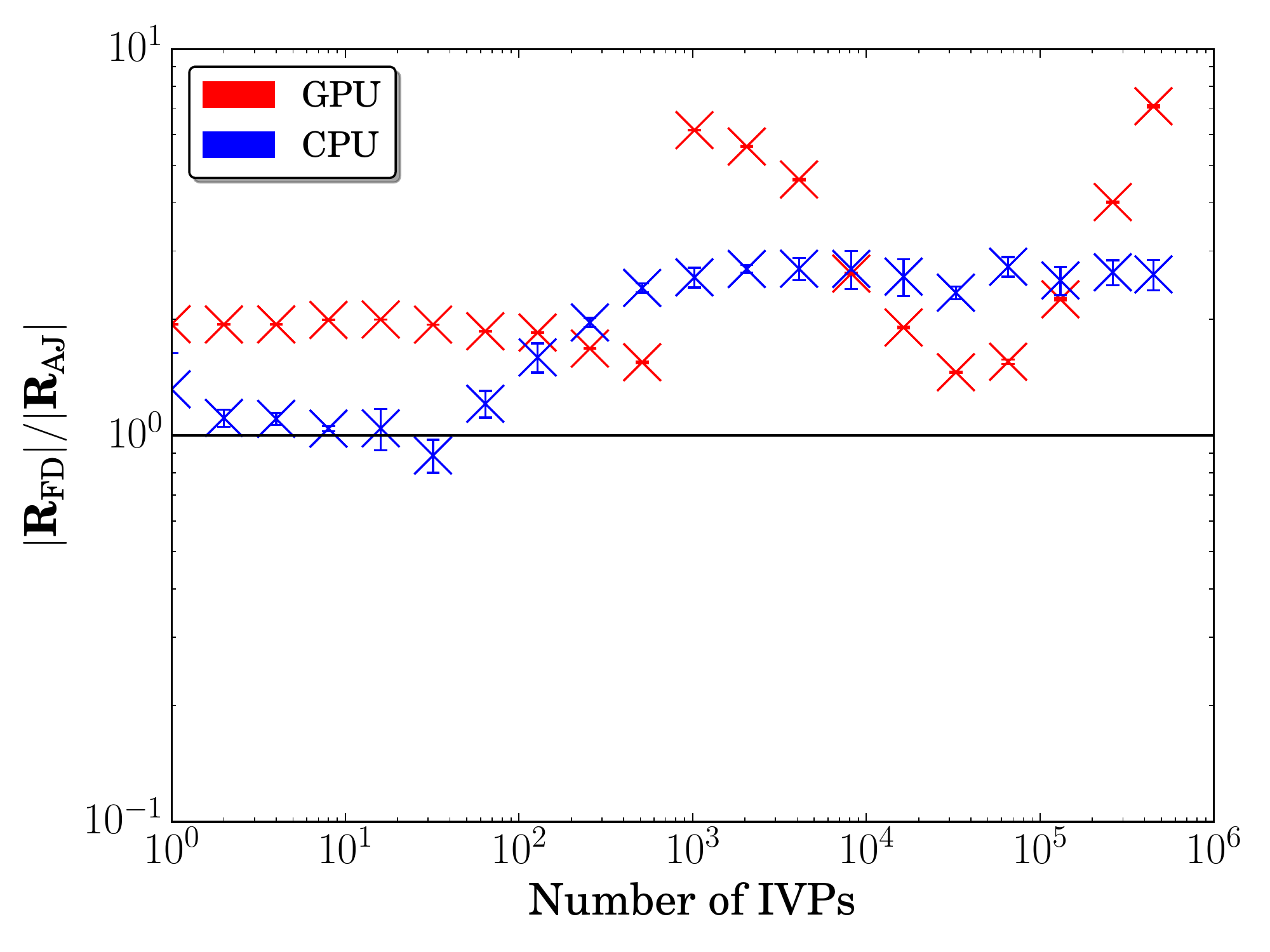}
      \caption{GRI-Mech 3.0 model with $\Delta t = \SI{1e-6}{\second}$}
      \label{F:AJ_ch4_small}
  \end{subfigure}
  \caption{Ratio of the average finite-difference Jacobian based integrator runtime $\lvert\textbf{R}_{\text{FD}}\rvert$ to that of the analytical Jacobian runtime $\lvert\textbf{R}_{\text{AJ}}\rvert$ for the \texttt{Radau-IIA} (CPU\slash GPU) solvers.
  Error bars indicate standard deviation, and the horizontal lines show a ratio of one.
  Data, plotting scripts, and figure files are available under CC-BY~\cite{paperscript:2017}.}
  \label{F:AJ_comp}
\end{figure}

Figure~\ref{F:AJ_comp} shows the speedup achieved on both the CPU and GPU for the \texttt{Radau-IIA} algorithm for various cases; the GRI-Mech 3.0 results for the larger global time-step size have been omitted due to long run times.
For the hydrogen model (Figs.~\ref{F:AJ_h2_small} and~\ref{F:AJ_h2_large}), using the analytical Jacobian offers minimal performance benefit for the CPU-based integrators, reaching a maximum speedup of \SI{1.49}{$\times$} and \SI{1.39}{$\times$} for the small and large global time-step sizes, respectively.
Our previous work~\cite{Niemeyer:2016aa} demonstrated that evaluation of the analytical Jacobian was \SI{5.28}{$\times$} faster on the CPU for the same chemical kinetic model; thus, the minor speedup seen here results from reuse of the Jacobian within the \texttt{Radau-IIA} solver, such that integration only requires a few Jacobian evaluations.
In some cases the finite-difference Jacobian solver may be faster than the analytical Jacobian solver; although it is difficult to explain the exact cause of this phenomena, differences in the finite-difference Jacobian likely caused the integrator to follow a slightly different instruction path (e.g., with fewer Jacobian updates\slash chemical source term evaluations) changing the integration cost.
However, for large numbers of conditions, the analytical-Jacobian-based CPU solver indeed performs faster than the finite-difference counterpart.
In contrast, the analytical-Jacobian-based GPU solver performs significantly faster than the finite-difference GPU solver in all cases for the hydrogen model, reaching a maximum speedup of \SI{12.16}{$\times$} for the smaller global time-step size.
As discussed in Sec.~\ref{S:divergence}, significantly higher levels of thread divergence are expected for the larger global time-step size.
Correspondingly, the maximum speedup of the GPU solver increases to \SI{240.96}{$\times$} for the larger global time-step size.
Figure~\ref{F:AJ_ch4_small} shows that the speedup of the CPU and GPU solvers reach \SI{2.61}{$\times$} and \SI{7.11}{$\times$}, respectively, for the larger GRI-Mech 3.0 model at the smaller global time-step size.
It is clear that for a per-thread-based GPU integrator, using an analytical Jacobian is essential for efficient integration due to thread-divergence concerns.

%

\section{Conclusions}
\label{S:conclusions}

The large size and stiffness of chemical kinetic models for fuels traditionally requires the use of high-order implicit integrators for efficient solutions.
Past work showed orders-of-magnitude speedups for solution of nonstiff to moderately stiff chemical kinetic systems using explicit solvers on GPUs~\cite{Niemeyer:2011aa,Le2013596,Niemeyer:2014aa}.
In contrast, work on stiff chemical kinetics integration with implicit GPU solvers has been limited to specialized cases, or failed to surpass current CPU-based techniques.

This work demonstrated and compared the performances of CPU- and GPU-based integration methods capable of handling greater stiffness, including an implicit fifth-order Runge--Kutta algorithm and two fourth-order exponential integration algorithms, using chemical source term and analytical Jacobian subroutines provided by the \texttt{pyJac} software~\cite{niemeyer_2016_51139,Niemeyer:2015ws,Niemeyer:2016aa}.
By comparing the performance of these algorithms using two chemical kinetic models, including hydrogen with 13 species and 54 reactions~\cite{Burke:2011fh} and methane with 53 species and 325 reactions~\cite{smith_gri-mech_30}, and using two global time-step sizes (\SI{e-6}{\second} and \SI{e-4}{\second}), we drew the following conclusions, q:
\begin{itemize}
 \item For global time-step sizes relevant to large-eddy simulations (e.g., $\Delta t = \SI{e-6}{\s}$), the GPU-based implicit Runge--Kutta method was roughly equivalent to the CPU-based implicit \texttt{CVODE} integrator running on \numrange{12}{38} CPU cores.
 \item At larger global time-step sizes, the performances of all GPU-based integrators decreased significantly due to thread divergence.
 \item For a global time-step size relevant to Reynolds-averaged Navier--Stokes simulations (e.g., $\Delta t = \SI{e-4}{\s}$), the GPU-based implicit Runga--Kutta solver performed equivalent to \texttt{CVODE} running on \num{15} cores for the hydrogen model, and just \num{3} cores for the GRI-Mech 3.0 model.
 \item The higher memory traffic required due to the size of the GRI-Mech 3.0 model significantly decreased GPU solver performance; a sparse analytical chemical kinetic Jacobian formulation must be developed to achieve high performance for still larger chemical kinetic models on the GPU.
 \item The exponential solvers were significantly less efficient than the implicit integrators on the CPU and GPU for all relevant cases.
 \item Using an analytical Jacobian matrix on the GPU is critical for efficient chemical kinetics integration due to thread divergence; speedups of \SIrange{7.11}{240.96}{$\times$} over a finite-difference-approximation were reached on the GPU, far surpassing the corresponding CPU speedup of \SIrange{1.39}{2.61}{$\times$}.
\end{itemize}

Based on these results, we conclude that the exponential solvers poorly fit the SIMT acceleration paradigm due to high levels of thread divergence combined with the relatively high cost of integration steps due to Arnoldi iteration (as compared with other semi-implicit integration techniques).
Instead, we recommend directing further focus on stiff semi-implicit solvers such as (non-exponential) Rosenbrock solvers, explored for the CPU by Stone and Bisetti~\cite{stone2014comparison}, and inexact Jacobian W-methods~\cite{steihaug1979attempt,Schmitt2004}.
Further improvements to the analytical Jacobian code, e.g., by using a chemical kinetic system based on species concentrations to increase Jacobian sparsity, are likely to further increase performance of the developed algorithms.
Additionally, newer GPUs should be tested to examine the ability of larger cache sizes and more available registers to improve performance by reduction of slow global memory loads\slash stores; a per-block solution still may need to be adopted for efficient integration of larger chemical kinetic models.
However, this work also showed that thread divergence poses a challenge to high performance of GPU-based integration techniques on a per-thread basis.
Our future work will therefore include a more comprehensive study of thread divergence, as well as developing methods to mitigate or eliminate its negative performance impact.
Finally, new integration techniques will be investigated and paired with work studying the selection of appropriate solvers based on estimated stiffness.

\section*{Acknowledgments}

This material is based upon work supported by the National Science Foundation under grants ACI-1534688 and ACI-1535065.

\appendix
\setcounter{figure}{0}

\renewcommand*{\thesection}{\appendixname~\Alph{section}}

\section{}
\label{S:supp}

The results for this paper were obtained using \texttt{accelerInt} v1.0-beta~\cite{accelerInt:beta}.
The most recent version of \texttt{accelerInt} can be found at its GitHub
repository \url{https://github.com/SLACKHA/accelerInt}.
All figures as well as the data and plotting scripts necessary to reproduce them, are
available openly under the CC-BY license~\cite{paperscript:2017}.

Supplementary material associated with this article includes unscaled
plots of integrator runtimes and characterizations of the partially stirred
reactor data for this work.

\clearpage

\bibliography{GPU-integrator-paper}
\bibliographystyle{elsarticle-num}

\end{document}